\documentclass[twocolumn]{aa}
\usepackage{txfonts}
\usepackage{graphicx}
\usepackage{subfigure}
\usepackage{float}
\usepackage{psfrag}
\input epsf
\newif\ifAMStwofonts
\def\wisk#1{\ifmmode{#1}\else{$#1$}\fi}

\def\nwm2sr {\wisk{\rm nW/m^2/sr\ }}
\def\nw2m4sr {\wisk{\rm nW^2/m^4/sr\ }}

\begin{document}
   \title{CMB Observations and the Production of \\ Chemical Elements at
   the End of the Dark Ages}

   \author{K. Basu, 
          \inst{1}
	  C. Hern\'andez--Monteagudo
	  \inst{1}
          \and
          R. Sunyaev\inst{1,2}
          }

   \offprints{K.Basu,\\ e-mail: kaustuv@mpa-garching.mpg.de}

   \institute{Max-Planck Institut f\"ur Astrophysik, 
    Karl-Schwarzschild-Str. 1, D-85740 Garching, Germany 
         \and
             Space Research Institute (IKI), 
 Profsoyuznaya 84/32, Moscow117810, Russia\\
             }

\date{Received ............. /  
      Accepted ............. }

   \abstract{
The metallicity evolution and ionization history 
of the universe must leave its imprint on 
the Cosmic Microwave Background  
through resonant scattering of CMB photons by atoms, ions and 
molecules. These transitions partially erase original temperature
anisotropies of the CMB, and also generate new fluctuations.
 In this paper we propose a method to
determine the abundance of these heavy species in low density 
(over-densities less than $10^4-10^5$) 
optically thin regions of the universe by using the unprecedented 
sensitivity of current and future CMB experiments. In particular,
we focus our analysis on the sensitivity of the PLANCK HFI 
detectors in four spectral bands. We also present 
results for l=220 and 810 which are of interest for balloon and
ground-based instruments, like ACT, APEX and SPT. We use the 
fine-structure transitions of atoms and ions as a source of 
frequency dependent optical depth ($\tau_{\nu}$). These transitions give
different contributions to the power spectrum of  CMB 
in different observing channels. By comparing 
results from those channels, it is possible to {\it avoid} the limit imposed
by the cosmic variance and 
to extract information about the abundance of corresponding species at the
redshift of scattering.  For PLANCK HFI we will be able to get strong 
constraints ($10^{-4}-10^{-2}$ solar fraction) on the abundances of
neutral atoms like C, O, Si, S, and Fe in the redshift 
range 1-50.
Fine-structure transitions of ions like CII, NII or OIII  set similar 
limits in the very important redshift range 3-25 and can be used to
probe the ionization history of the universe.
Foregrounds and other
frequency dependent contaminants may set a serious limitation 
for this method. 
   \keywords{cosmology: cosmic microwave background -- cosmology:
theory -- intergalactic medium -- atomic processes -- abundances}
  
} 

   \authorrunning{K.Basu, C.Hern\'andez-Monteagudo and R.Sunyaev}
   \titlerunning{CMB Observations \& Production of Chemical Elements}
   \maketitle
%

\section{Introduction}

High precision observations of CMB anisotropies are giving us
unique information about the angular
distribution of CMB fluctuations, as well as their spectral dependence 
in a very broad frequency range.
HFI and LFI detectors of PLANCK\footnote{PLANCK URL site: 
http://astro.estec.esa.nl/Planck/} spacecraft will provide
unprecedented sensitivity in 9 broad band $( \Delta \nu / \nu \sim
20-30 \% )$ channels, uniformly distributed in the spectral region of the
CMB where contribution of different foregrounds are expected to be
at a minimum. CMBpol and other proposed missions are
expected to reach noise levels 20 - 100 times lower than that of
PLANCK HFI with technology already available (Church 2002).
 The WMAP\footnote{WMAP URL site: http://map.gsfc.nasa.gov/} satellite
was designed to provide measurements of the CMB temperature anisotropies
in the whole sky with an average sensitivity of 35$\mu $K per 
$0.3^{\circ} \times 
0.3^{\circ}$ pixel at the end of the mission (Bennett et al. 2002, Page et
al. 2002). Once such sensitivity limit is reached,
the WMAP data will be the first real-life test for all those attempts to 
estimate,
with extremely high precision, the key
parameters of our universe using this sensitivity
(Bond, Efstathiou \& Tegmark 1997, Einsestein, Hu \& Tegmark 1999,
Prunet, Sethi \& Bouchet, 2000). Indeed, after the first 12 months of 
operation, the WMAP team is recovering the first multipoles of the CMB power
 spectrum with an accuracy of a few percent (Hinshaw et al. 2003). 

 In this paper we are presenting an additional use of the tremendous
sensitivity of PLANCK and CMBpol, and ground-based experiments like
ACT, APEX and SPT.  
We propose to look for or to place upper limits on the abundances of
heavy elements present in the inter galactic medium and/or in
 optically thin clouds of gas everywhere in the redshift range $[1,500]$.
 We shall focus on the fine-structure lines of neutral 
(CI, OI, SiI, FeI, ...) and ionized (CII, NII, OIII, ...) atoms, 
which might provide information about the epoch of first star
formation and ionization history of the universe.
Limits on abundances of heavy atoms and ions 
 can be obtained by utilizing the frequency dependent
opacity that will be generated by scattering of background photons by
these species; giving rise to different $\delta
C_l$-s in in different observing bands of the experiments.
Although Planck HFI is expected to detect the combined signal from the
strongest lines, like OI $63.2 \mu$, CII $157.7 \mu$ and OIII $88.4
\mu$, future multichannel broadband CMB anisotropy experiments 
like CMBpol might
permit to detect contributions from these lines separately 
 from the epoch when dark ages were terminating, universe became 
partially ionized and heavy element production begun. It is important
 that the polarization signal arising due to resonance scattering 
depends strongly on the properties of the 
 transition (Sazonov et al. 2002), and together with 
the temperature signal will permit us to
separate contributions of different species.  

The observed primordial acoustic peaks and angular fluctuations 
should not depend on the frequency at all. This is connected
with the nature of Thomson scattering which produces these
fluctuations both in the time of recombination of hydrogen in the
universe (Peebles \& Yu 1970, Sunyaev \& Zel'dovich 1970) and during the
secondary ionization of the universe (Sunyaev 1977, Ostriker \& Vishniac 1986,
Vishniac 1987, and more recently Gruzinov \& Hu 1998, da Silva et al. 2000,
Seljak, Burwell \& Pen 2000, Springel, White \& Hernquist 2001, and 
Gnedin \& Jaffe 2001). In this context, WMAP polarization measurements have 
recently shown
strong evidence favouring an early reionization scenario, with  
$z_r = 20^{+10}_{-9}$ (Kogut et al. 2003, 95\% confidence). But 
if any amount of chemical elements are present during the
dark ages, then these species will be able to scatter the CMB in their
fine-structure lines. This scattering
would not only partially smooth out primordial CMB anisotropies, but 
in addition will generate
new fluctuations  through the Doppler shift of the line
associated with the motion of matter connected with the growth of density
perturbations. The main difference with Thompson scattering is that
the latter is giving us equal contribution over the whole CMB 
spectrum, whereas the
discussed line scattering would give different contribution to
different observing channels placed at different parts of the
spectrum.  Likewise, the contribution in this case would be 
restricted to a very {\em thin} slice in the universe.
Hence there is a possibility to
detect contribution from the lines with high transition
probabilities, even though the typical optical depth we would find
is very small ($\tau_{\nu}< 10^{-4}$).  Since 
every line is able to work only in a given range of redshift,
 having observations on different wavelengths can give an
upper limit to abundances of different species at different epochs.  
This method relies critically in the fact that
intrinsic CMB temperature fluctuations are frequency independent. Therefore,
in the absence of other frequency-dependent components, the difference of
two CMB maps obtained at different frequencies must be sensitive to the
difference in abundance of resonant species at those redshifts probed by the
frequency channels. For this reason, this method is particularly sensitive
to the possible presence of any frequency-dependent signal coming from 
foregrounds.  

The idea to use the line transitions of atoms and molecules for
modifying the CMB power spectrum is not new. 
Dubrovich (1977, 1993) proposed the use of rotational lines of primordial
molecules (LiH, HD etc.) as a source of creating new angular fluctuations in
the CMB, and there have been attempts to observe these molecules at
high redshifts (de Bernardis et al. 1993).
 Varshalovich, Khersonskii and Sunyaev
(1978) were trying to estimate the spectral distortions of CMB due to
absorption in fine-structure lines of oxygen and carbon at redshifts
around 150-300 where the temperature of electrons should be lower than
the temperature of CMB due to different adiabatic indices of radiation
and matter (see Zel'dovich, Kurt \& Sunyaev 1968). 
Suginohara et al.(1999)
probed the possibility of detecting excess flux due to emission in these
lines coming from very over-dense regions in the universe.
  Recently Loeb (2001) and Zaldarriaga \& Loeb (2002) (hereafter ZL02)
computed the distortions connected with the recombination of
primordial Lithium and scattering in the Lithium resonant doublet
line. Unfortunately the wavelength of fine-structure 2P-2S transition
of Lithium atoms is too short and will be unobservable by PLANCK and
balloon instruments.  
On the very low frequency domain, future experiments like SKA and
LOFAR might detect signal from neutral hydrogen 21 cm line, which also
carry important ioformation from high redshifts (Sunyaev \& Zel'dovich
1975, Madau, Meiksin \& Rees 1997). 
We consider below the ground-state fine-structure
lines of heavier elements, with wavelengths of the order of $50-200 \mu$,  
 which will in principle be observable with
PLANCK and CMBpol 
if they are present in the redshift range $[1,500]$.  The problem of
overcoming their extremely small optical depth comprises the main idea
of this paper.

We will not discuss in detail the origin
and ways of enrichment of the inter-galactic gas by heavy elements.
This certainly requires existence of massive stars, supernova
explosions, stellar and galactic winds, and even jets 
from disks around young stars 
with cold molecular gas  
(Yu, Billawala \& Bally 1999). The main goal of
this paper is to show that the announced sensitivity of PLANCK
detectors might permit us to set very strong upper limits to the time
of enrichment of inter-galactic gas by heavy elements, the time of
reionization, and maybe even to detect the heavy elements in
the inter-cluster medium. 
The census of baryons in the local universe (Fukugita et al. 1998) 
shows that most of the baryon remains unobserved, and the proposed method
might set way to detect its existence at high redshifts, when it had
moderate or low temperature. These missing baryons are centainly
out of stars, interstellar gas and intergalactic gas in clusters
and groups of galaxies. However, we know that such baryons
should exist because they have been detected by WMAP at the 
last scattering surface at
$z \simeq 1100$, and are also necessary to justify the observed
abundance of deuterium and $^6$Li in the early universe.
Due to the
wide range of redshifts under study, and the uncertain degree of
mixture and clumpiness of our species in the interstellar and
inter-galactic  
medium, we shall assume that all elements 
are smoothly distributed in the sky. In other words, in the present
paper we shall address exclusively the homogeneous low density 
optically thin case, where effects
related to strong over-density 
of gas and collisional excitation are excluded
and left as subject of an upcoming work. Resonant scattering effects
will produce the discussed signal even if all the gas in the enriched
plasma has over-density up to $10^3$. Under  this 
{\it smooth} approximation, we will
find that the effect of resonant species on the CMB power spectrum
will be particularly simple, especially in the high multipole range.
Indeed, in the small angular scales, we shall show that the induced
change in the $C_l$'s is given by $\delta C_l \simeq -2\;\tau_{X_i} C_l$,
where $\tau_{X_i}$ is the optical depth induced by the resonant scattering
of species $X$ and $C_l$ is the primordial CMB power spectrum, which is
currently the main target of most CMB experiments.

In section 2 we describe the basic approach for computing 
optical depths and corresponding deviations
 in the CMB power spectra due to scattering, and then    
show how we can put constraints on abundances using the
sensitivity of present and future CMB experiments.   
Section 3 presents our main results for atoms and ions, 
and in section 4 we discuss various possible ionization and enrichment
scenarios of the universe and the nature of predicted signal for each
case. We will demonstrate in the Appendix B that optically thin
over-dense regions with $\langle n_e \rangle < 10-200$ cm$^{-3}$ are
contributing to the effect of resonant scattering. 
The effect of foreground emissions on our analysis is discussed in
section 5, and we conclude in section 6.  
Throughout this paper we have used the
parameters for standard $\Lambda \mbox{CDM}$ cosmology 
(Spergel et al. 2003),  with
$\mbox{H}_0 = 71 \ km \ s^{-1} \ Mpc^{-1}, \ \ \Omega_b = 0.044, \ \
\Omega_m = 0.27, \ \Omega_{\Lambda} = 0.73, 
\ \mbox{and} \ \tau_T = 0.17$ for secondary reionization.  
$\mbox{T}_0$ is the present-day value of mean CMB
temperature, given by $\mbox{T}_0 = 2.726 \mbox{K}$ (Mather et
al. 1994). \\


\section[]{Basic Approach}

We now discuss the basic method of obtaining the deviations in the CMB
power spectrum by resonance scattering of atoms and ions, and
using this deviation to constrain their abundance during the dark
ages. The interaction of CMB photons with atoms 
and ions will mostly consist of resonant scattering with either
an atomic or a rotational/vibrational transition, 
depending on the species under study.
In the Appendix A we detail how  we introduce the optical depth due 
to the resonant transition ($\tau_{X_i}$)
in the CMBFAST  code (Seljak \& Zaldarriaga, 1996). 
We consider a particular resonant transition $i$ for a given species
$X$, with rest-frame resonant frequency $\nu_i$. 
The total optical depth encountered by CMB photons on their way from
the last scattering surface to us is then obtained by adding the
contributions from all lines to the standard Thomson opacity:   
\begin{equation}
\tau = \tau_{T} + \sum_i \tau_{X_i}
\label{eq:tautot}
\end{equation}

In order to calculate
$\tau_{X_i}$, we shall recur to the formula which gives the optical 
depth of a resonant transition in an expanding medium (Sobolev 1946),

\begin{equation}
\tau_{X_i}(z) = f_i \ {\frac{\pi e^2}{m_e c}} {\frac{\lambda_i
n_{X_i}(z)}{H(z)}},
\label{basic-eqn}
\end{equation}

where  $f_i$ is the absorption oscillator strength of the resonant transition,
$\lambda_i$ is the corresponding wavelength (in rest frame), 
$n_{X_i}(z)$ is number density of $X$ species at redshift $z$, 
and $H(z)$ is the Hubble parameter at that epoch. 
The oscillator strength depends on $\lambda_i$, the Einstein coefficient of the
transition $A_{ul}$, and the degeneracy of the levels involved in it:
\begin{equation}
f_{i} = \frac{m_e c}{8 \pi^2 e^2}  
 \frac{g_u}{g_l} \   \lambda_i^2  A_{ul}
\label{oscillator}
\end{equation}

A simple and elegant treatment following Gunn \& Peterson (1965) 
using $\delta$-function line profile, instead of thermally broadened
gaussian  
for the same line, gives same value of optical depth.
Further modifications are made to this formula to take
into account the finite populations in the upper transition levels,
which is important for atomic fine-structure transitions, 
where excitation temperature for them is
comparable to the radiation temperature at high redshifts. 
Also the effect of non-zero cosmological constant can not be neglected
in the low redshift universe of interest.  
Considering these facts, 
we obtain the following formula for optical depths: 

\[
\tau_{X_i}(z)  \ = \ 
 1.7 \times 10^{-6} \
  \left( \frac{{\cal X}_{X_i}(z)}{10^{-6}} \right)  
 \ \ \times \]
\begin{equation}
\phantom{xxxxxxxxxxxx}
  \times \ 
  \left( {\frac{{\cal S}(z)}{{\cal S}(z=10)}} \right) \ 
   \left( \frac{\lambda_i}{100 \ \mu} \right) 
  \ \left( \frac{f_i}{10^{-9}} \right) \ {\cal B}. 
\label{sobolev}
\end{equation}

Here ${\cal X}_{X_i}(z)$ is the ratio between the number density
$N_{X_i}(z)$ of the atomic or ionic species $X$ under consideration,
with respect to the baryon number density at the same redshift:
$N_b(z) = 2 \times 10^{-7}\; (1+z)^3$ cm$^{-3}$. i.e. ${\cal
X}_{X_i}(z) = N_{X_i} / N_b$ presents the evolution of abundance of
the given species due to element production, ionization and
recombination processes.   
We propose to constrain the minimum
abundance that can be detected at that redshift in {\it solar units},
or $[X]_{min} \equiv {\cal X}_{X_i}(z)\; / \;{\cal X}_{\odot}$, where,
for example, ${\cal X}_{\odot} = 3.7 \times 10^{-4}$ for carbon. 
${\cal S}(z)$ gives the redshift dependence of optical depth
in a $\Lambda$CDM universe: ${\cal S}(z) = (1+z)^3 \  
[(1+z)^2 (1+\Omega_m z) - z (2+z) \Omega_{\Lambda}]^{-
\frac{1}{2}} $ (see, e.g., Bergstrom 1998)
, and $\lambda_i$ is the wavelength in micron. The final term   
${\cal B}$   accounts for the 
actual fraction of atom/ion present in their ground state, 
and is governed solely by the temperature of background 
radiation at the redshift of scattering. For a two-level system, this
fraction is simply 
\begin{equation}
n_l = [1+(g_u/g_l)\ \exp(-h \nu_i/kT_0(1+z))]^{-1}.
\label{terror}
\end{equation} 
We also include here the correcting term for
the induced emission in the presence of the CMB and finite population
of the upper level:   
\begin{equation}
  {\cal B} = n_l \ \times \ 
\left[ 1 - \exp{\left( -\frac{h\nu_i}{k_B T_0 (1+z)} \right)}
\right].
\label{eq:Boltz}
\end{equation}

The resonance scattering on ions and atoms in thermal equilibrium 
with black body radiation does not change its intensity. However, 
the observed CMB also has finite primordial angular fluctuations. 
The effect of resonant scattering is to decrease
these angular fluctuations, and to bring the system more
close to thermodynamic equilibrium. Therefore 
resulting fluctuations observed on the frequency of the 
line should differ from the situation on other 
frequencies which are far from the resonance.

At low multipoles peculiar motions arising due to the growth of 
large scale density perturbations become important. All ions or 
atoms are moving in the same direction and change the frequency of
CMB photons during resonant scattering. This leads to
generation of new anisotropies of background.

\medskip
We next give a simple description of the modification of the power spectrum
of the CMB when it encounters a resonant species.
As mentioned in the Introduction, 
our first hypothesis will be that the species responsible for the scattering
are homogeneous, isotropic and smoothly  (i.e., not clumped) 
distributed in the Universe, at least
during the epoch of interaction with the CMB. 
One may argue that heavy species are located in halos where the 
first stars are produced and that their distribution in the sky can non be 
regarded as {\it smooth}. In this case, the final effect would depend on the
typical angular size of the patches in which the species have been spread and
on their total sky coverage. This paper will observe the case
where those scattering patches percolate in the sky, 
giving rise to a smooth, 
homogeneous picture. The case of patchy distribution of emitting 
sources will be addressed in a forthcoming work, where collissional 
processes are studied in an extremely dense optically thick environment.

In the conformal Newtonian gauge,
(also known as the longitudinal gauge), the {\it k}-mode of the photon
temperature fluctuation at current epoch is given by (Hu \& Sugiyama
1994) 
\[
\Delta_T (k,\eta_0,\mu ) \  = \ 
\int_0^{\eta_0}d\eta \;e^{ ik\mu(\eta-\eta_0) }
\bigl[ \Upsilon (\eta ) \left( \Delta_{T0} 
 - i\mu v_b \right) 
	 \] 
\begin{equation}
\phantom{xxxxxxxxxxxxxxxxxxxxxxxxxxxxxxxx}
 + \ \dot{\phi} + \psi  - ik\mu\psi \bigr].
\label{eq:dT1}
\end{equation}

Here we have neglected the polarization term (which contributes at 
most with a 
few percent of the temperature amplitude). $\eta $ denotes  
the conformal time, $\Upsilon \equiv {\dot \tau}(\eta)\exp{-\tau (\eta )} $ 
is the visibility function  and
$\tau(\eta) = \int_{\eta}^{\eta_0} d\eta' \dot{\tau} (\eta')$ is the 
optical depth.
 $\Delta_{T0}$ accounts for the intrinsic temperature fluctuations,
$v_b$ for the velocity of baryons and $\phi$ and $\psi$ are the scalar 
perturbations of the metric in this gauge. We have also neglected all
tensor perturbations. If we now introduce the optical depth associated to a
resonant transition as a Dirac delta of amplitude $\tau_{X_i}$ placed at 
$\eta=\eta_{X_i}$, 
\begin{equation}
 d\tau(\eta) = \sigma_T n_e(\eta) d\eta + \tau_{X_i} 
\delta(\eta-\eta_{X_i}) d\eta,
\label{eq:smplapp}
\end{equation} 
we readily obtain that the original anisotropies have been erased by a
factor $e^{-\tau_{X_i}}$, whereas new anisotropies have been generated at 
the same place:
\begin{equation}
\Delta_T (k,\eta_0,\mu )\;  = \;e^{-\tau_{X_i}}\Delta_{T_{orig}}
		 (k,\eta_{X_i},	\mu )\; + \;
	\Delta_{T_{new}} (k,\eta_{X_i},\mu )
\label{eq:newdT}
\end{equation}

 In real space, if we assume that $\tau_{X_i}\ll 1$ , 
the Dirac delta approximation for the resonant transition
(eq.(\ref{eq:smplapp})) translates into 
\[
\frac{\delta T}{T_0} ({\bf n},z=0) = \left(1-\tau_{X_i}\right)\;
	\frac{\delta T}{T_0}
({\bf n},z_{X_i} )
\]
\begin{equation}
\phantom{xxxxxxxxxxx}
+ \tau_{X_i}\;
\frac{\delta T}{T_0}\biggr|_{new}^{lin}({\bf n},z_{X_i})\; +  \;
{\cal O}\left[\tau_{X_i}^2\right].
\label{eq:newdT2}
\end{equation}
Here, ${\bf n}$ is the unitary vector denoting the observing direction, 
and $\delta T/T_0|_{new}^{lin}({\bf n},z_{X_i})$ is the
linear in $\tau_{X_i}$ term of the newly generated anisotropies\footnote{ 
In what follows, our temporal coordinate will be denoted by redshift
or conformal time $\eta$ indiferently.}. We remark again
that we are assuming that the species is homogeneously distributed,
so that $\tau_{X_i}$ does not depend on position.
Let us now consider two different
observing frequencies: the first one corresponds to the 
scattering redshift $z_{X_i}$, and the another corresponds to a
scattering redshift too high to expect any significant presence of
the species $X$. If 
we compute the correlation function, $\bigl< \frac{\delta T}{T_0} ({\bf
n_1},z=0) \cdot  \frac{\delta T}{T_0} ({\bf
n_2},z=0)\bigr>$, in both cases, we find that the difference of these
quantities will be equal to:\footnote{This
type of manipulations of the correlation function/power spectrum which
avoids the Cosmic Variance limit is
possible as long as the weak signal (signal induced by scattering in this
case) has different
spectral dependence than the CMB.}
\[
\Delta \left( \bigl< \frac{\delta T}{T_0} ({\bf
n_1},z=0) \cdot  \frac{\delta T}{T_0} ({\bf
n_2},z=0)\bigr>\right) = 
\]
\[
\phantom{xxxxxxx}
\tau_{X_i} \biggl( -\;2\;\bigl< \frac{\delta T}{T_0}
({\bf n_1},z_{X_i}) \cdot \frac{\delta T}{T_0}
({\bf n_2},z_{X_i}) \bigr> \; 
\]
\begin{equation}
\phantom{xxxxxxx}
+\; 2\; \bigl< \frac{\delta T}{T_0}
({\bf n_1},z_{X_i}) \cdot \frac{\delta T}{T_0}\biggr|_{new}^{lin}({\bf
n_2},z_{X_i})\bigr>\biggr) + {\cal O}\left[\tau_{X_i}^2\right].
\label{eq:dCf}
\end{equation}
That is, the term linear in $\tau_{X_i}$ of $\Delta 
\left( \bigl< \frac{\delta T}{T_0} ({\bf
n_1},z=0) \cdot  \frac{\delta T}{T_0} ({\bf
n_2},z=0)\bigr> \right)$ is the sum of two
different contributions: the suppression of
original fluctuations (which, as we shall see, dominates at small angular
scales), and the cross-correlation of the newly generated anisotropies
with the intrinsic CMB field.  
The first term (suppression) does not depend on the potential or velocity
fields during scattering, but only on the intrinsic CMB anisotropy field. 
 Moreover, we must 
remark that those contributions are evaluated {\em at scattering
epoch}\footnote{The dependence on cosmic epoch of eqs.
(\ref{eq:newdT2},\ref{eq:dCf}) has been simplified. 
We again refer to 
Appendix A for a formal derivation.}. 
Hence, if the resonant scattering takes place {\em before}
reionization, the changes in the correlation function (or in the power
spectrum, as we show below) will be sensitive to the CMB anisotropy field
{\em before} being processed by the {\it re-}ionized medium.
In the Appendix A we give a detailed computation of the change in the power
spectrum due to a resonant line. 
As we concentrate in the optically thin limit, it is
possible to make a power expansion on the optical depth $\tau_{X_i}$:
\begin{equation}
\delta C_l \equiv C_l^{X_i} - C_l = 
		\tau_{X_i} \cdot {\cal C}_1 + \tau_{X_i}^2 \cdot {\cal C}_2 +
			{\cal O}( \tau_{X_i}^3 ).
\label{eq:deltaCl}
\end{equation}

In this equation, $C_l^{X_i}$ and $C_l$ refer to the angular power spectrum
multipoles in the presence and in the absence of the $X_i$ resonant
transition, respectively. 
For the limit of very small $\tau_{X_i}$ one can retain only the linear order,
and a direct identification of $\delta C_l$ with the abundance $[X](z)$ is
possible by means of eq.(\ref{sobolev}): the change induced in the CMB 
power spectrum will be proportional to the abundance of the species
 responsible for the resonant scattering. 
 In agreement with what 
has been established when studying $\Delta \left( 
\bigl< \frac{\delta T}{T_0} ({\bf
n_1},z=0) \cdot\frac{\delta T}{T_0} ({\bf
n_2},z=0)\bigr> \right)$, we find that 
${\cal C}_1$ changes sign at some multipole $l_{cross}$: positive values
of $\delta C_l$ imply generation of new anisotropies ($l<l_{cross}$),
whereas anisotropies are suppressed for $l>l_{cross}$ (see figure
\ref{oxygen2} at very low $l$). 
In the Appendix
A we note that positive values of $\delta C_l$ (generation) are 
due to the non-zero correlation existing between anisotropies generated
during recombination and those
generated during
the resonant scattering. This correlation is due to the coupling of $k$ modes
of the initial metric perturbation field in scales of the order or
larger than the distance separating the two events (recombination 
at redshift $1100$ and resonant scattering at redshift $5$-$25$),
 (Hern\'andez--Monteagudo \& Sunyaev, in preparation). 

\begin{figure}
\centering
\psfrag{123456789l}{Multipole, $l$}
\psfrag{123456789123456789cl}{$ l(l+1)/2 \pi \ C_l \ \ (\mu K^2)$}
\psfrag{123456789log}{log($[\mbox{O}]_{min} / [\mbox{O}]_{\odot}$)}
\includegraphics[width=10.5cm, height=7cm]{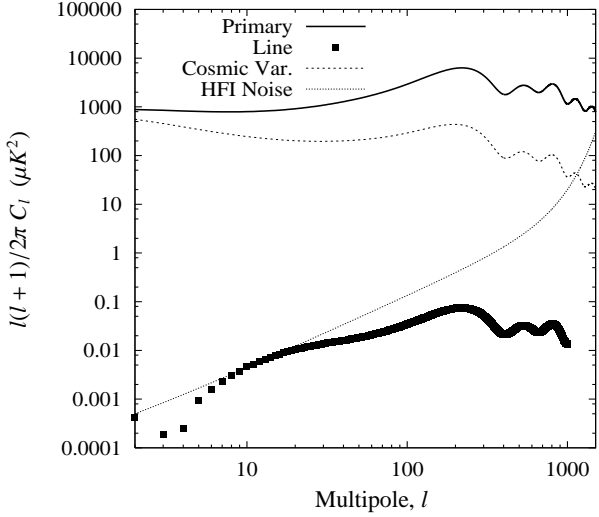}
\caption{The nature of temperature anisotropy that will be 
marginally detectable
with Planck HFI 143 GHz channel, using $63 \mu$ OI line. 
Abundance is taken from the table(\ref{hfitable}),  
as $5.3 \times 10^{-4}$ 
solar at redshift 32.
Here {\bf Primary} denotes  the measured temperature anisotropy ({\it
upper line}), $C_l^{pri} + \delta C_l \approx C_l^{pri}$, 
and {\bf Line} denotes the
the newly generated anistropy, $|\delta C_l|$, arising from
line scattering ({\it bottom, with filled squares}).
This $|\delta C_l|$ is obtained by
taking the difference from the 100 GHz channel, and 
changes sign around  $l=3$. 
The corresponding noise level is denoted by {\bf HFI noise}.  
Also shown is the cosmic variance limit 
({\it short-dashed line}) for comparison.}
\label{oxygen2}
\end{figure}

\subsection{$\delta C_l$'s at small angular scales}

\begin{figure}
\centering
\includegraphics[width=7.5cm, height=6cm]{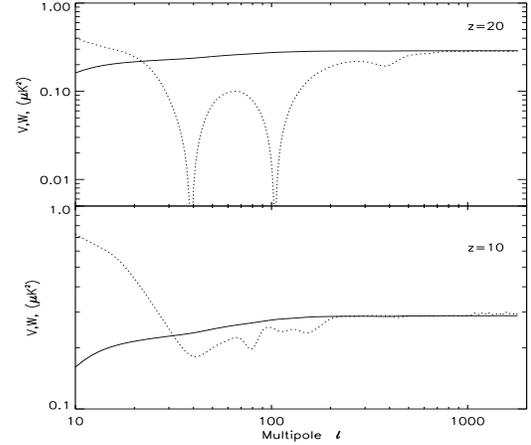}

\caption[columnwidth]{For two lines placed at different redshifts, 
($z=20$ in the top,
$z=10$ in the bottom), we plot the quantities $V$ (solid lines) and 
$W$ (dotted lines), defined as
$V \equiv \left|  C_l^{reio} / C_l  -1 \right|$ and 
$W \equiv \left| \left| \delta C_l^{reio} / \delta C_l \right| -1 \right|$,
 respectively.
In both cases, in the limit of high $l$, these quantities tend to
$2\; \tau_{reio}$, where $\tau_{reio}$ was taken to be $0.17$. In both
panels, $\tau_{X_i}=10^{-3}$.}

\label{fig:dlta_reio}
\end{figure}

With respect to the high-l range, we find that the change induced in the
power spectrum takes a very simple form:

\begin{equation}
\delta C_l \simeq - 2\; \tau_{X_i}\; C_l
\label{simple}
\end{equation}

where $C_l$ is the intrinsic power spectrum. 
This dependence is identical to the effect of reionization
on the power spectrum at small angular scales. Indeed, in that scenario, if
the optical depth due to electron scattering during this epoch is given
by $\tau_{reio}$, then we have that for $l\gg 1$ the intrinsic CMB power
spectrum generated at recombination is suppressed by a factor 
$1-\exp({-2\tau_{reio}})$, or $\approx -2 \tau_{reio}$ if $\tau_{reio} < 1$. 
So in both (resonant and Thompson) scatterings, the shape of the
induced change in the power spectrum is particularly simple and equal to
$\delta C_l \simeq -2\tau\;C_l$, for a given optical depth $\tau$
and intrinsic power spectrum $C_l$. This simplifies considerably
the effect of reionization on the $\delta C_l$'s induced by resonant
transitions. Indeed, if the
symbol $\Delta^{reio}$ denotes the difference of a given quantity evaluated
in the presence and in the absence of reionization, then we have that, for
high $l$, $\Delta^{reio}( C_l) = -2\tau_{reio}\;C_l$ and 
$\Delta^{reio} \left(\delta C_l \right) = -2\tau_{reio}\; \left( \delta C_l
\right)$. This is shown in figure (\ref{fig:dlta_reio}), where we plot
the quantities $V \equiv \left| C_l^{reio} / C_l  -1 \right|$ and
$W \equiv \left| \left| \delta C_l^{reio} / \delta C_l \right| -1 \right|$, 
that is, the
relative change of $C_l$ and $\delta C_l$ due to the presence of
reionization. We have taken $\tau_{reio}=0.17$. We see that, for high $l$,
both $V$ and $W$ approach the limit $V \simeq W \simeq 2\;\tau_{reio}$, i.e.,
the effect of reionization is identical in the two cases.

\subsection{Measuring $\delta C_l$'s and abundances}

Next we investigate the limits of detectability of the
optical depth 
of atoms and ionss in current and future CMB missions.
Our starting point is the expected uncertainty in the 
obtained $C_l$'s from any CMB experiment (Knox 1995):
\begin{equation}
\sigma_{C_l}^2 = \frac{2}{\left(2l+1\right)f_{sky}}
	        \left( C_l + w^{-1}B_l^{-2} \right)^2.
\label{eq:senst1}
\end{equation}
In this equation, 
$C_l$ is the underlying CMB power spectrum, $f_{sky}$ is the fraction of the
sky covered by the experiment, and $w$ is the pixel weight given by
$w^{-1}=\sigma_N^2 \Omega_{pix}$, with $\sigma_N$ the noise amplitude for
pixels of solid angle $\Omega_{pix}$.  $B_l$ is the beam
window function, and we shall approximate it by a gaussian.
We are assuming that the noise is gaussian, uniform and uncorrelated. 
The first term
in parentheses of eq.(\ref{eq:senst1}) 
reflects the uncertainty associated to the cosmic variance.
In the ideal case of an experiment with no noise, this would be the 
unavoidable
uncertainty when identifying the {\it observed} $C_l$'s to the $C_l$'s 
corresponding to a particular cosmological model. However, our interest
focuses on the comparison of power spectra measured in different frequency
 channels with respect to a {\it measured} reference power spectrum,
(which is supposed to be free of any {\it contamining} species of atoms and/or
ions). Under the assumption that the reference power spectrum
contains only CMB and noise components, we can write the following expression
for the uncertainty in the measured power spectrum difference between
a {\it probe} channel and the reference channel:
\[
	 \sigma_{ \delta C_l }^2 = 
	\frac{2}{(2l+1)f_{sky}}  
			\bigl[ \left( \delta C_l^{prob} + 
                       w^{-1}_{prob} B_{l,prob}^{-2} \right)^2 \]
\begin{equation}
\phantom{xxxxxxxxxxxxxxxxxxxxxxxxxxxxx}
		 + \left( w^{-1}_{ref} B_{l,ref}^{-2} \right)^2 \bigr].
\label{eq:newsgcl}
\end{equation}
The indexes {\it prob} and {\it ref} refer to the {\it probe} and 
{\it reference} 
channel respectively, and the first term $\delta C_l^{prob}$ 
refers to the cosmic
variance associated to the temperature anisotropies generated
by the resonant species. 

Let us now assume that a CMB experiment is observing at a frequency for which
it is expected to see the effects of a resonant transition of a species $X$
at redshift $1+z_{X_i}= \nu_{X_i}/\nu_{obs}$. One can then make use of 
eq.(\ref{eq:deltaCl}) to relate the minimum $\tau_{X_i}$ observable with the 
sensitivity of the experiment:
\begin{equation}
\left( \tau_{X_i} \right)_{min} \simeq 
        \;n \;\frac{ \sigma_{\delta C_l} }{{\cal C}_1}.
\label{tauminn}
\end{equation}
The factor $n$ expresses the $\sigma$-level necessary to claim a
detection. For all the results presented in this paper, we have
set the detection threshold at 3$\sigma$, i.e. $n$ = 3. 
Fig.\ref{oxygen2} shows how a very low optical depth suffices to
generate changes in the power spectrum (filled squares) 
which are big enough
to overcome the combined noise of the two experiment channels being
compared, (dotted line). Once this
minimum optical depth is obtained, we can easily find the
corresponding minimum abundance through eq.(\ref{sobolev}), which is
one of the main goals of our paper.  
In fact, this limit can be further improved by a factor of 
$\sim \sqrt{\Delta l}$ if one computes band power spectra on some 
multipole range $\Delta l$.
We shall address the details of these issues in the next section, where 
we show that for PLANCK's HFI channels, 
values for the optical depth as low as 
$10^{-5}-10^{-7}$ can be detected. \\

So far we are neglecting the problematic 
associated to the calibration between channels and the possible presence of 
foregrounds. Given the frequency dependent nature of the latter, 
the amplitudes and their characterization of the different galactic and 
extragalactic contaminants are critical for our purposes. 
We return to these issues in section 5, were we estimate the effect of
foreground contamination on our analysis.\\

\section{Results and Discussion}

Now we present the results for selected atoms and ions and that
can produce measurable distortions in the CMB spectrum. 
As mentioned previously, the focus will be on the fine-structure
transitions for atoms and ions
, as their frequencies fall in the far-infrared and microwave
range and are therefore perfectly suited to our purpose.

We divide this section into two parts: the first part shows results
for all the important atomic and ionic species. 
In second part we briefly mention  
the contribution from the ovedense regions to the effect considered. 

\subsection{Scattering by atoms and ions of heavy elements}


\begin{table*}
\centering

\begin{tabular}{ccccccccc} \hline
 {Atom/} &
 {Wavelength} & {Oscillator} & {HFI freq.} &
{Scattering} & {$\cal B$} & {Opt. depth for} 
 & {$[X]_{min}$ for} & {$\langle [X]_{min} \rangle$ in}   \\ 

 {Ion} & {(in $\mu$)} & {strength} &
 {(GHz)} & {redshift} & {factor} & {$10^{-2}$ solar abundance} 
  & {$l=10$} & {\  $l = 10-20$ \ }  \\ 
  \hline\hline

{\bf C I} & 609.70  & $ 1.33 \times 10^{-9} $ & 143 & 2.4 & 0.76 
        & $6.4 \times 10^{-6}$  
        & $5.3 \times 10^{-3}$ & $2.6 \times 10^{-3}$  \\ 

       &   &  & 217 & 1.3 & 0.92   
        & $3.9 \times 10^{-6}$  
        & $1.4 \times 10^{-2}$ & $6.8 \times 10^{-3}$ \\

       &   &  & 353 & 0.4 & 0.99  
        & $1.6 \times 10^{-6}$  
        & $2.1 \times 10^{-1}$ & $1.2 \times 10^{-1}$  \\

  \cline{2-9}

        &   370.37 & $ 9.08 \times 10^{-10} $ & 143 & 4.7 & 0.15
        & $1.2 \times 10^{-6}$ 
        & $2.8 \times 10^{-2}$ & $1.3 \times 10^{-2}$  \\ 

       &   &  & 217 & 2.8 & 0.09  
        & $3.7 \times 10^{-7}$  
        & $1.6 \times 10^{-1}$  & $8.1 \times 10^{-2}$  \\      

   \hline

{\bf C II} & 157.74  & $ 1.71 \times 10^{-9} $ & 143 & 12.3 & 0.79 
        & $1.8 \times 10^{-5}$ 
        & $2.7 \times 10^{-2}$ & $6.2 \times 10^{-3}$  \\ 

       &   &  & 217 & 7.9 & 0.94   
        & $1.1 \times 10^{-5}$  
        & $7.7 \times 10^{-3}$ & $3.0 \times 10^{-3}$   \\

       &   &  & 353 & 4.4 & 0.99  
        & $5.6 \times 10^{-6}$  
        & $7.7 \times 10^{-2}$ & $3.6 \times 10^{-2}$   \\

   \hline

{\bf N II}  & 205.30 & $ 3.92 \times 10^{-9} $ & 143 & 9.2 & 0.76
        & $1.1 \times 10^{-5}$ 
        & $7.6 \times 10^{-3}$ & $2.6 \times 10^{-3}$  \\ 

       &   &  & 217 & 5.8 & 0.92  
        & $6.8 \times 10^{-6}$  
        & $8.6 \times 10^{-3}$ & $3.8 \times 10^{-3}$   \\

       &   &  & 353 & 3.1 & 0.99  
        & $3.5 \times 10^{-6}$  
        & $1.3 \times 10^{-1}$ & $6.8 \times 10^{-2}$   \\

  \cline{2-9}

      &    121.80  & $ 2.74 \times 10^{-9} $ & 143 & 16.2 & 0.16
        & $2.1 \times 10^{-6}$  
        & $1.3 \times 10^{-1}$ & $3.8 \times 10^{-2}$  \\ 

      &     &  & 217 & 10.5 & 0.09 
        & $6.4 \times 10^{-7}$  
        & $3.4 \times 10^{-1}$ & $1.1 \times 10^{-1}$   \\

   \hline

{\bf N III} & 57.32 & $ 4.72 \times 10^{-9} $ & 143 & 35.6 & 0.79 
        & $2.5 \times 10^{-5}$ 
        & $2.3 \times 10^{-3}$ & $7.4 \times 10^{-4}$  \\

      &     &  & 217 & 23.4 & 0.94 
        & $1.5 \times 10^{-5}$  
        & $6.1 \times 10^{-3}$ & $2.0 \times 10^{-3}$   \\


   \hline

{\bf O I} & 63.18  & $ 3.20 \times 10^{-9} $ & 143 & 32.2 & 0.88
        & $1.0 \times 10^{-4}$  
        & $5.3 \times 10^{-4}$ & $1.7 \times 10^{-4}$  \\ 

      &    &  & 217 & 21.2 & 0.96 
        & $6.3 \times 10^{-5}$  
        & $2.0 \times 10^{-3}$ & $6.4 \times 10^{-4}$  \\

      &    &  & 353 & 12.5 & 1.00 
        & $3.1  \times 10^{-5}$  
        & $2.2 \times 10^{-1}$ & $4.9 \times 10^{-2}$   \\

%

  \hline

{\bf O III}  & 88.36 & $ 9.16 \times 10^{-9} $ & 143 & 22.8 & 0.76
        & $2.2 \times 10^{-4}$  
        & $3.5 \times 10^{-4}$ & $1.2 \times 10^{-4}$   \\ 

      &    &  & 217 & 14.8 & 0.92 
        & $1.4 \times 10^{-4}$  
        & $8.4 \times 10^{-3}$ & $1.8 \times 10^{-3}$   \\

      &    &  & 353 & 8.6 & 0.99 
        & $7.4 \times 10^{-5}$  
        & $1.2 \times 10^{-2}$ & $4.4 \times 10^{-3}$   \\

  \cline{2-9}

          & 51.81 & $ 6.55 \times 10^{-9} $ & 143 & 39.5 & 0.17 
        & $4.5 \times 10^{-5}$   
        & $1.4 \times 10^{-3}$ & $4.7 \times 10^{-4}$   \\ 

          &    &   & 217 & 26.0 & 0.10  
        & $1.4 \times 10^{-5}$  
        & $6.5 \times 10^{-3}$ & $2.2 \times 10^{-3}$   \\

   \hline

{\bf Si I} & 129.68  & $ 6.24 \times 10^{-9} $ & 143 & 15.2 & 0.76
        & $6.4 \times 10^{-6}$ 
        & $6.9 \times 10^{-2}$ & $1.8 \times 10^{-2}$  \\ 

       &    &  & 217 & 9.8 & 0.92  
        & $4.2 \times 10^{-6}$  
        & $3.7 \times 10^{-2}$ & $1.2 \times 10^{-2}$  \\

  \cline{2-9}

          & 68.47  & $ 4.92 \times 10^{-9} $ & 143 & 29.7 & 0.19 
        & $1.8 \times 10^{-6}$   
        & $3.1 \times 10^{-2}$ & $1.0 \times 10^{-2}$  \\ 

   \hline

{\bf Si II} & 34.81  & $ 7.74 \times 10^{-9} $ & 217 & 39.2 & 0.94
        & $2.0 \times 10^{-5}$  
        & $4.8 \times 10^{-3}$ & $1.6 \times 10^{-3}$  \\

       &   &  & 353 & 23.4 & 0.99 
        & $1.0 \times 10^{-5}$  
        & $7.4 \times 10^{-3}$ & $2.4 \times 10^{-2}$  \\

   \hline

{\bf S I} & 25.25  & $ 8.03 \times 10^{-9} $ & 217 & 54.4 & 0.96  
        & $6.0 \times 10^{-6}$  
        & $2.4 \times 10^{-2}$ & $7.4 \times 10^{-3}$   \\ 

   \hline

{\bf Fe I} & 24.04  & $ 1.69 \times 10^{-8} $  & 143 & 86.4 & 0.83
        & $3.2 \times 10^{-5}$  
        & $4.9 \times 10^{-3}$ & $1.3 \times 10^{-3}$  \\

       &   &  & 217 & 57.2 & 0.95  
        & $2.0 \times 10^{-5}$  
        & $7.7 \times 10^{-3}$ & $2.3 \times 10^{-3}$  \\

       &   &  & 353 & 34.4 & 0.99  
        & $9.7 \times 10^{-6}$  
        & $7.5 \times 10^{-2}$ & $2.5 \times 10^{-2}$  \\

   \cline{2-9}

           & 34.71  & $ 2.06 \times 10^{-8} $ & 143 & 59.5 & 0.02 
        & $9.2 \times 10^{-7}$  
        & $1.1 \times 10^{-1}$ & $3.4 \times 10^{-2}$  \\  

   \hline

{\bf Fe II} & 25.99  & $ 1.73 \times 10^{-8} $ & 217 & 52.9 & 0.95
        & $1.9 \times 10^{-5}$  
        & $7.1 \times 10^{-3}$ & $2.2 \times 10^{-3}$  \\ 

           &  &  & 353 & 31.7 & 0.99  
        & $1.3 \times 10^{-7}$  
        & $7.4 \times 10^{-2}$ & $2.4 \times 10^{-2}$  \\ 

   \hline

{\bf Fe III} & 22.93  & $ 3.09 \times 10^{-8} $ & 353 & 36.1 & 0.99
        & $1.8 \times 10^{-5}$  
        & $4.2 \times 10^{-2}$ & $1.4 \times 10^{-2}$  \\

   \hline

\end{tabular}

\caption{Minimum abundance of the most important 
atoms and ions that can be detected from Planck
HFI. We have used a fixed value of abundance (1\% solar) to obtain
the optical depth in the fine-structure transitions of any given
species, and because of the very low values of optical depths, have
used the linear relation from eqn.(\ref{tauminn}) to obtain the
minimum detectable abundance for the sensitivity limit of Planck
HFI. The $\cal B$ factor is the correction term obtained from
eqn.(\ref{eq:Boltz}). In the last column we have further improved the
sensitivity by averaging the instrument noise in the multipole range
$l=10-20$. 
The 100 GHz channel has been used as 
reference for all the cases. 
We present central redshift for corresponding channel, however,  
in reality HFI will be 
 able to give limits only for redshift intervals
corresponding to the widths of frequency channels.}

\label{hfitable}
\end{table*}


In this section we investigate the possibility of distorting the CMB
power spectrum by scattering from neutral atoms like CI, OI, SiI, SI,
FeI etc., as well as
singly and doubly ionized species of
heavy elements, like CII, NII, SiII, FeII, OIII etc.  This is
important for various reasons.  
The Gunn-Peterson effect permits us to prove that the universe was
completely ionized as early as redshift 6, up to the position of most
distant quasars known today.  Recent results from WMAP satellite
pushes the reionization redshift as far as $z=20$, suggesting a
complex ionization history (Kogut et al, 2003).
There is extensive discussion about the nature of reionization, and
also on the possibility for universe being reionized twice (Cen 2003). 
  In any case, before the universe was ionized completely,
there were regions of ionized medium around first bright stars and
quasars.  It will be difficult to prove that the universe was
partially ionized using Ly-$\alpha$ line because of its extremely high
oscillator strength, which makes the universe optically thick in this
line even when the neutral fraction of Hydrogen is only $10^{-6}$  at
$z = 6$.  However, the infrared lines that we are discussing in this paper
have much weaker oscillator strength and therefore neutral gas is
transparent in these lines up to very high redshifts, even if we assume
solar abundance.  This fact permits us to consider the possibility 
that Planck and other future CMB experiments setting very strong 
limits on abundance of neutral atoms in the early universe.

The effect of line emission to the thermal spectrum
of CMB can be estimated in a simple order-of-magnitude way by the
formulation given in Appendix B.  
If there were significantly over-dense regions at high redshifts (up to $z
\sim 20$) which were completely ionized and enriched with metal ions
(e.g. C$^+$, N$^+$, O$++$, Fe$^+$ etc.), 
then collisional excitation followed by radiative de-excitation 
will be a significant source of emission in the same fine-structure
lines.
However, to make line emission visible we need three factors:
high abundance of the elements, high density of the electrons in the 
strongly over-dense regions and large amount of over-dense
regions in the volume which we are investigating. 
Hence it is much more promising to study angular distortions
of the CMB generated by scattering from the low density regions of the
universe, rather than studying distortions in its thermal spectrum, 
for constraining heavy element abundances at high redshifts. Such low
density inter-galactic gas is believed to contain most of the baryonic
mass of the universe, and possibly exists as warm/hot gas with $10^5 <$
T$_e < 10^7$ K today (Cen \& Ostriker 1999). 
But at redshifts $z>1$ this inter-galactic gas
should have moderate (T$_e \sim 10^4$ K) 
or low (T$_e \sim$ T$_{CMB}(z)$) temperature, and the
proposed method of observing angular fluctuations caused by scattering
from neutral or singly ionized atoms might set a direct way to
detect its existence.

As mentioned above, 
the main contributors of opacities in the relevant frequency range 
are oxygen, nitrogen, carbon, sulfur, silicon, and iron, along with minor
contributions mainly from phosphorus, aluminum, chlorine and nickel.
The $63 \mu$ fine-structure line of neutral oxygen gives strong
constraint on neutral species at high redshift, but early reionization
makes lines of CII, NII and OIII even more important. 
All data relating to fine-structure lines have been taken from
the ISO line-list for far-IR spectroscopy (Lutz, 1998), and the Atomic
Data for the Analysis of Emission Lines by Pradhan and Peng
(1995). When necessary, this compilation was supplemented by 
the freely available NIST Atomic
Database\footnote
{http://physics.nist.gov/cgi-bin/AtData/lines\_form}. \\

The basic idea of obtaining limits on abundances 
can be understood from Fig.\ref{oxygen2} and 
table \ref{hfitable}.  
Each broad-band channel of HFI 
acts in a specific range of redshifts
for a particular line, and we have tabulated the central redshift
corresponding to the scattering for three most sensitive channels for
several atomic and ionic fine-structure transitions.  
The lowest frequency channel of 100 GHz  
is assumed to be
``clean'' from line scattering, and thus used as reference. 
 We use a fixed abundance ($10^{-2}$ solar) to obtain the optical
depth in accordance with formula (\ref{sobolev}).  
Such small values of optical depths
allow us to use the first-order approximation, and so we finally
obtain the minimum optical depth 
, and hence the minimum abundance ({\it last column}, with respect 
to solar value) from the sensitivity level of the detector 
(at $3\sigma$ level). We have neglected signals below $l=5$,
especially at the quadrupole or $l=2$ where noise level is minimum,
due to the fact that it will be very difficult to observe the
predicted signal at such large angular scales due to the
foregrounds. The best angular range for Planck HFI is $l \sim 10-30$,
and  we can average it over a multipole
range $\Delta l$ and improve the detectibility by a factor of
$\sqrt{\Delta l}$. This result is shown in the {\it last} column of
our table.

Fig.\ref{oxygen2} shows the expected behavior of $\delta C_l$-s
when neutral oxygen is 
marginally detectable with the 143 GHz channel of HFI,
limiting the oxygen abundance as low as $10^{-4}$ 
relative to solar at redshift 30.
We show both the measured temperature anisotropy and $\delta C_l$-s
generated by line scattering in this plot. This line contribution
touches the noise limit at around $l \sim 10$, showing the angular
scale where best possible limit can be obtained.  We see that this
effect always lies much below the cosmic variance limit, but due to
the frequency dependence of new anisotropies we are not constrained by
this limit. 

Part of the results of this paper are summarized in Table \ref{hfitable}. 
This lists all
the atoms and ions on whose abundances we can put strong
limits, and all these limits are computed at $3\sigma$ level. 
With HFI we have the possibility to use more than one probe
channel to show different upper limits for the same species at
different redshifts.   This fact can be helpful to model the 
abundance history of the universe, and we present a
general discussion in the next section.


\subsection{Contribution from over-dense regions}

Enrichment of primordial gas by heavy
elements occurs due to supernova explosions of the first stars. High
velocity stellar and galactic winds and low velocity jets from disks
around forming stars, and objects of the type of SS 433 with baryonic
jets carry enriched matter to a large distance from the forming
stars. The observation of the most distant galaxies and quasars are
showing that even most distant objects ($z \sim 5-6$) have chemical
abundance on the level of solar (Freudling at al. 2003, Dietrich et
al. 2003).  
At the same time there is a
possibility that the low density matter, e.g. in future voids, will
have extremely low abundance of heavy elements. Observational method
we are proposing might permit us to observe ions and atoms of heavy
elements in {\it diffuse matter}, with over-density lower than
$10^3-10^4$ at redshifts $10-20$, and up to $\delta \sim 10^5-10^6$ at
$z \sim 2-5$. This means that even Ly-$\alpha$ clouds are contributing
to our effect. We should be careful only with the most damped
Ly-$\alpha$ systems, and with dense gaseous nebulae of the type of
Orion and dense giant molecular clouds. Diffuse gas in the galaxies
and proto-galaxies should also contribute to our signal. In Appendix B
we estimated the level of densities in the gas clouds, when the discussed
effect is diminished by 30\% or 50\% in comparison with the case of
diffuse inter-galactic space, and showed that over-densities greater
than $10^3$ are needed at high redshifts before collisional effect
begins to decrease amplitude of scattering signal. 
To summarize, the proposed method might
permit us to get signal from all diffuse matter of the universe,
excluding only the extremely over-dense, or optically thick clouds. 
It is probable that the over-dense regions were the first 
to be enriched by heavy
elements. Therefore it is important that the moderately over-dense 
regions of the universe are also contributing to the resonance scattering 
signal in the CMB angular fluctuations.

The simultaneous effect of free-free, line and dust emission 
from non-uniformly distributed over-dense regions at high redshifts,
which are in non-linear stage of evolution and entering the state of
intense star formation, produces an independent signal from the
scattering effect considered in this paper. 
The same lines which contribute to the $C_l$-s
due to resonant scattering from extended low density regions, would also
contribute to the power spectrum at smaller angular scales 
due to emission from over-dense regions. 
But over-dense objects like damped Ly-$\alpha$ absorbing systems and  
low density Ly-$\alpha$ clouds,    
together with stars, atomic and molecular gas in galaxies and hot gas in
clusters, contains only 20\% - 40\% of the
baryons in the universe (Fukugita et al. 1998, 
Penton et al. 2000, Valageas et al. 2001); meaning that 
resonance scattering from low density, optically thin gas with
low temperature and moderate ($<$$10^3$-$10^5$, see Appendix B)
over-density 
will always create its own distortion in the CMB
power spectrum alongside the emission generated from denser regions.  
Both effects carry information about the abundances of atoms and ions
averaged over the volume defined by multipole $l$ (or angular scale), 
and the frequency resolution of the detector which gives us 
the thickness of the slice in redshift space along line of sight. 
Resonant scattering effect
is sensitive only to the mean density of the scattering species  
$\langle n_{X_i} \rangle$ in that volume, whereas 
the line radiation effect is connected with collisions and therefore its
contribution to $C_l$'s depends on 
$\langle n_e n_{X_i} \rangle$ in the same volume. They are, thus, 
two independent effects that carry complementary information.
This latter effect is sensitive to the most over-dense regions in the 
universe, and will be discussed in detail in a subsequent work 
(Basu, Hern\'andez-Monteagudo and Sunyaev, in preparation).


\section {The Ionization History of The Universe}

Detection  of the OI, CII and OIII lines of the two most abundant 
elements will permit not only to trace the enrichment of the
universe by heavy metals, but also might open the way to follow
the ionization history of the universe.
According to recent models of stellar evolution very massive Pop III 
stars efficiently produce heavy metals like oxygen, carbon, silicon 
and sulfur (Heger \& Woosley, 2002). 
The CNO burning phase of the stars appearing immediately
afterwards will also produce large amount of nitrogen, and we can
expect strong signal from the ionized nitrogen $205 \mu$ line, 
as the time of
evolution of the first stars is extremely short ($\sim 10^6$ years) in
comparison with the Hubble time even at redshift $25$.

The WMAP finding that universe has rather high optical depth due
to secondary ionization at redshift $z_r = 20^{+10}_{-9}$
 (Kogut et al. 2003, 95\% confidence)
forced many theoretical groups to return to the picture of early
ionization due to Pop III stars (Cen 2003, Wyithe \& Loeb 2003). 
One of the possible evolution scenario of abundances for the elements
produced by these massive hot Pop III stars 
and intense star and galaxy formation is given in Fig.3a. Here we
consider two enrichment histories of the universe, with low ({\it A})
and high ({\it B}) metal abundances after reionization. The first
phase of metal enrichment occurs during the epoch of Pop III stars,
and at later epochs (redshift $3-5$) intense galaxy formation causes
further rise in the metallicity. Later in this section we consider a
third enrichment history with late reionization and metal injection.

\begin{figure*}
\centering
\includegraphics[width=10cm,height=10cm]{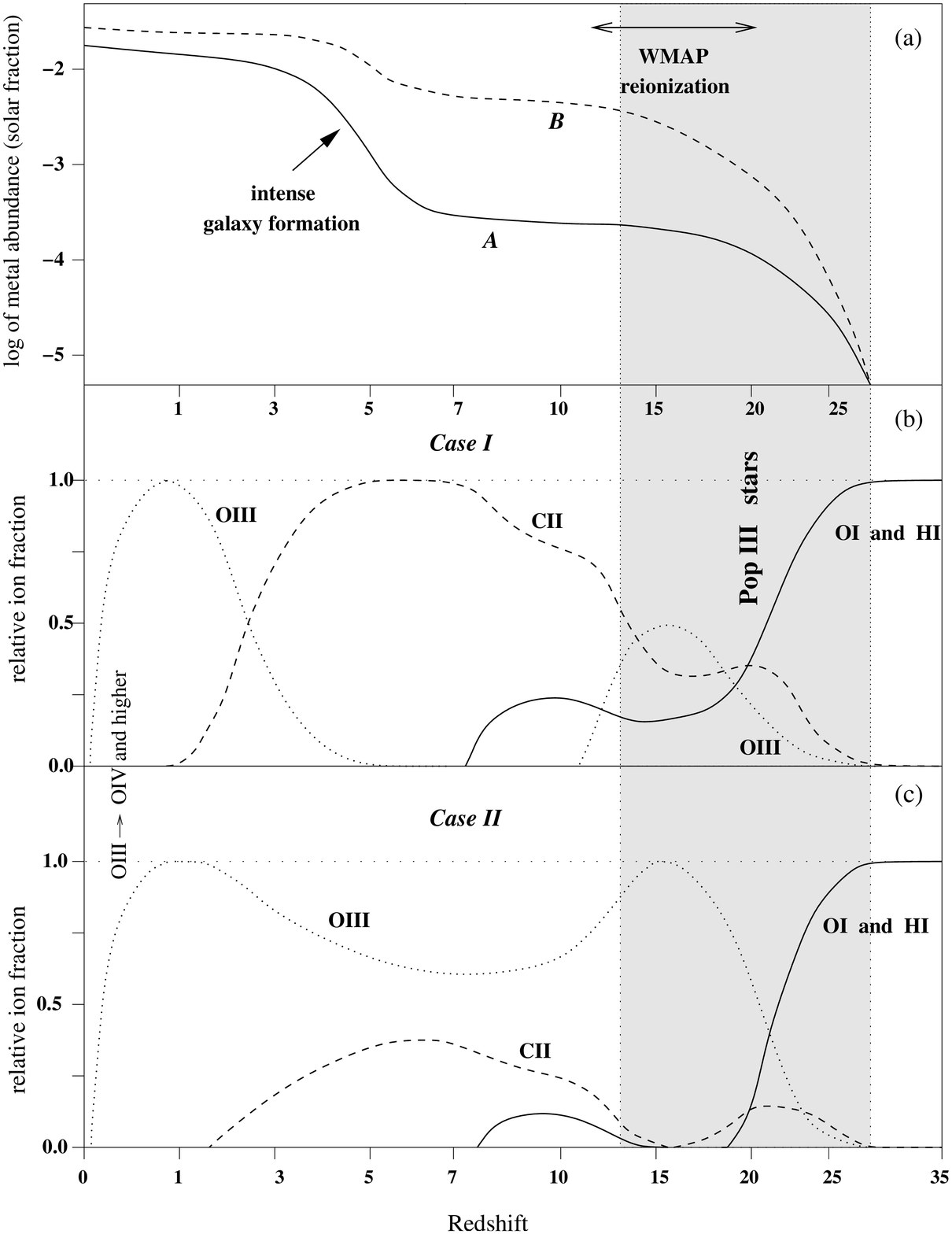}%

\vspace*{.5cm}

\includegraphics[width=8cm, height=5.5cm]{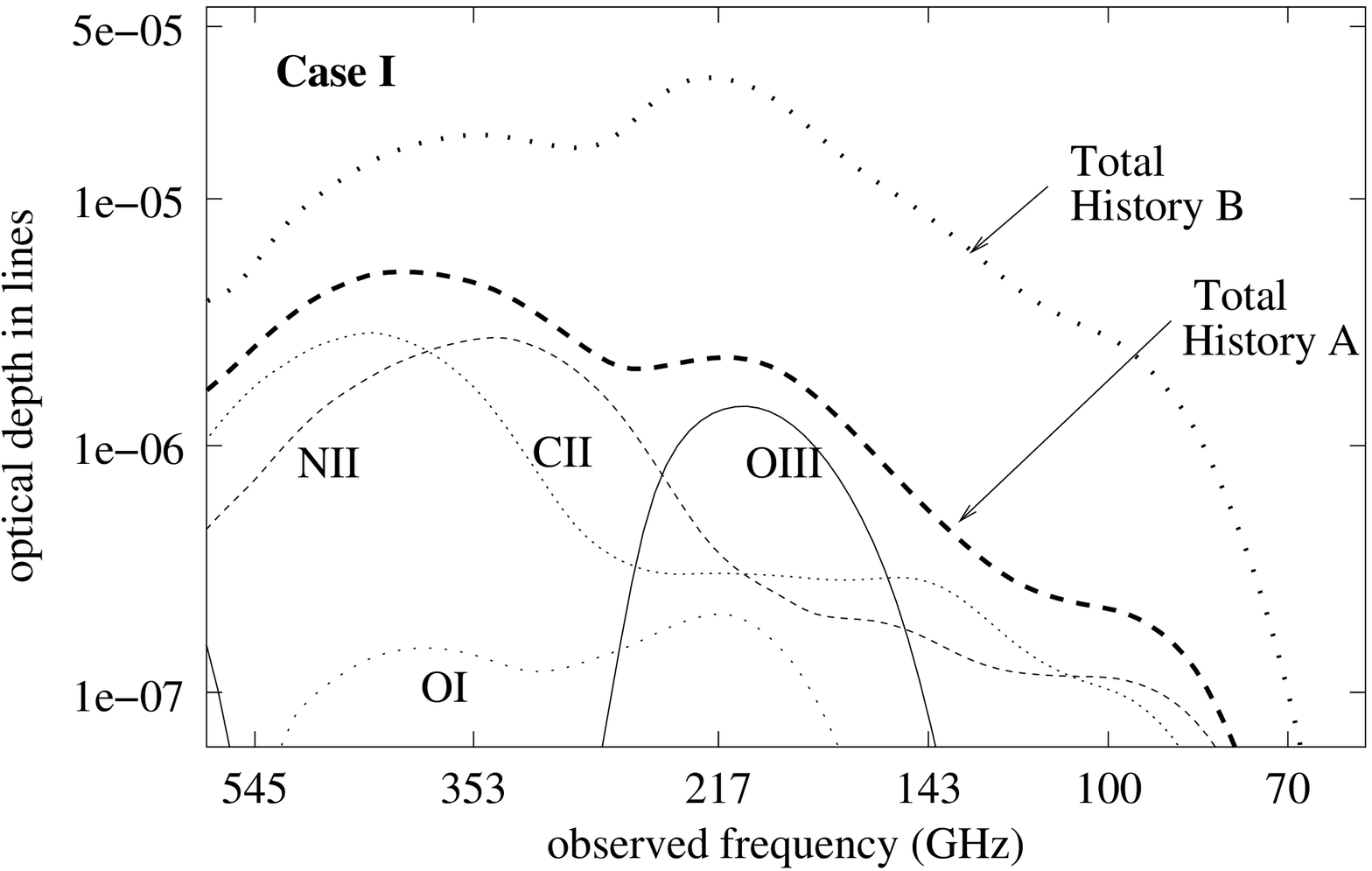}
\hspace*{.5cm}
\includegraphics[width=8cm, height=5.5cm]{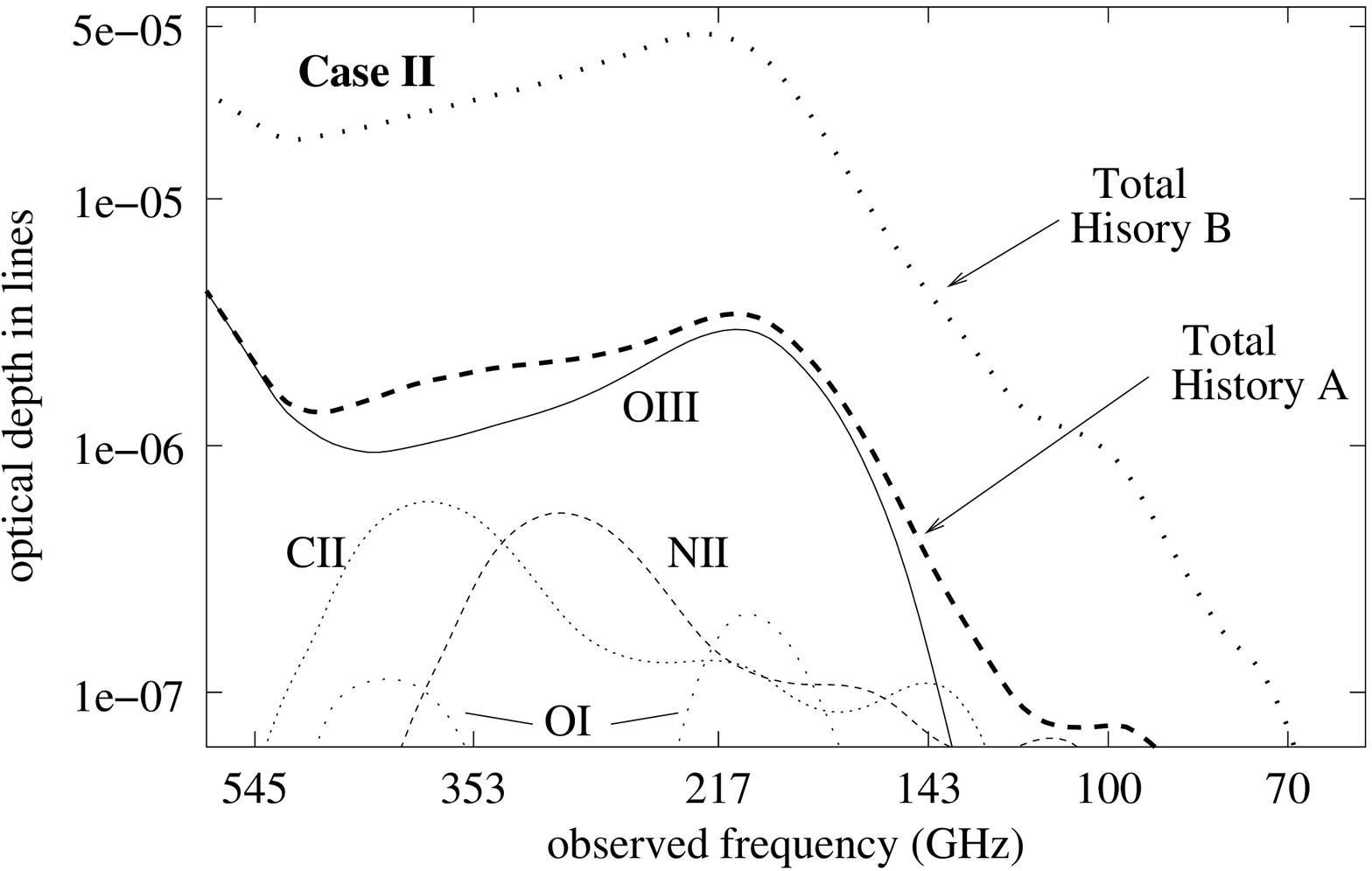}

\vspace*{.5cm}

\caption{ {\bf Top:} 
A schematic diagram for the abundance history of the
universe. ({\it a}) 
The upper panel shows the global pattern of metal abundances,
showing the two major epochs of enrichment of the IGM: first, during
the peak of activity of massive Pop III stars around redshift $15-25$
 ({\it shaded area}),
and second, during the peak of galaxy formation 
around redshift $3-5$ when global rate of
star formation reaches maximum. ({\it b})
Middle panel shows the relative fraction of three major atomic and
ionic species: OI (or HI), CII and OIII, normalized so that the total
abundance of all ions of a given element is close to unity at any
redshift. The OI abundance closely follows the neutral
hydrogen fraction of the universe because their almost similar
ionization potential (see discussion in
text).   
The redshift scale is chosen as $log (z+5)$ to emphasize the redshift
region $10-30$ of interest to this paper. 
At very low
redshifts ($ z < 0.7$) the IGM gets heated to very high temperatures
($T \sim 10^5-10^7$ K) causing even higher ionized species to exist, e.g.
OIII $\rightarrow$ OVI.
({\it c}) Lower panel shows another variation of relative ion
fraction, where Pop III stars ionize all the oxygen around redshift
15, so that OIII have higher abundance and correspondingly OI and CII
have lower abundance. 
{\bf Bottom:} Frequency dependence of optical depth in atomic and 
ionic fine-structure lines, 
in accordance with the abundance history sketched above. 
Shown here are the contributions from the four most important lines:
neutral oxygen $63 \mu$, doubly ionized oxygen $88 \mu$, singly
ionized nitrogen $205 \mu$ and singly ionized carbon $158 \mu$, 
and their total contribution for
each histories. History of NII ion is taken as similar to that of CII
ion. }
\label{skplot}
\end{figure*}

\begin{figure*}
\psfrag{tempanisotropy}{$l(l+1)/2\pi \ |\delta C_l|$ \ \ ($\mu$K$^2$)}
\psfrag{123456789empty}{}
\psfrag{obsfrequency}{$\nu_{obs}$ \ \ (GHz)}
\centering

\includegraphics[width=9cm, height=6.5cm]{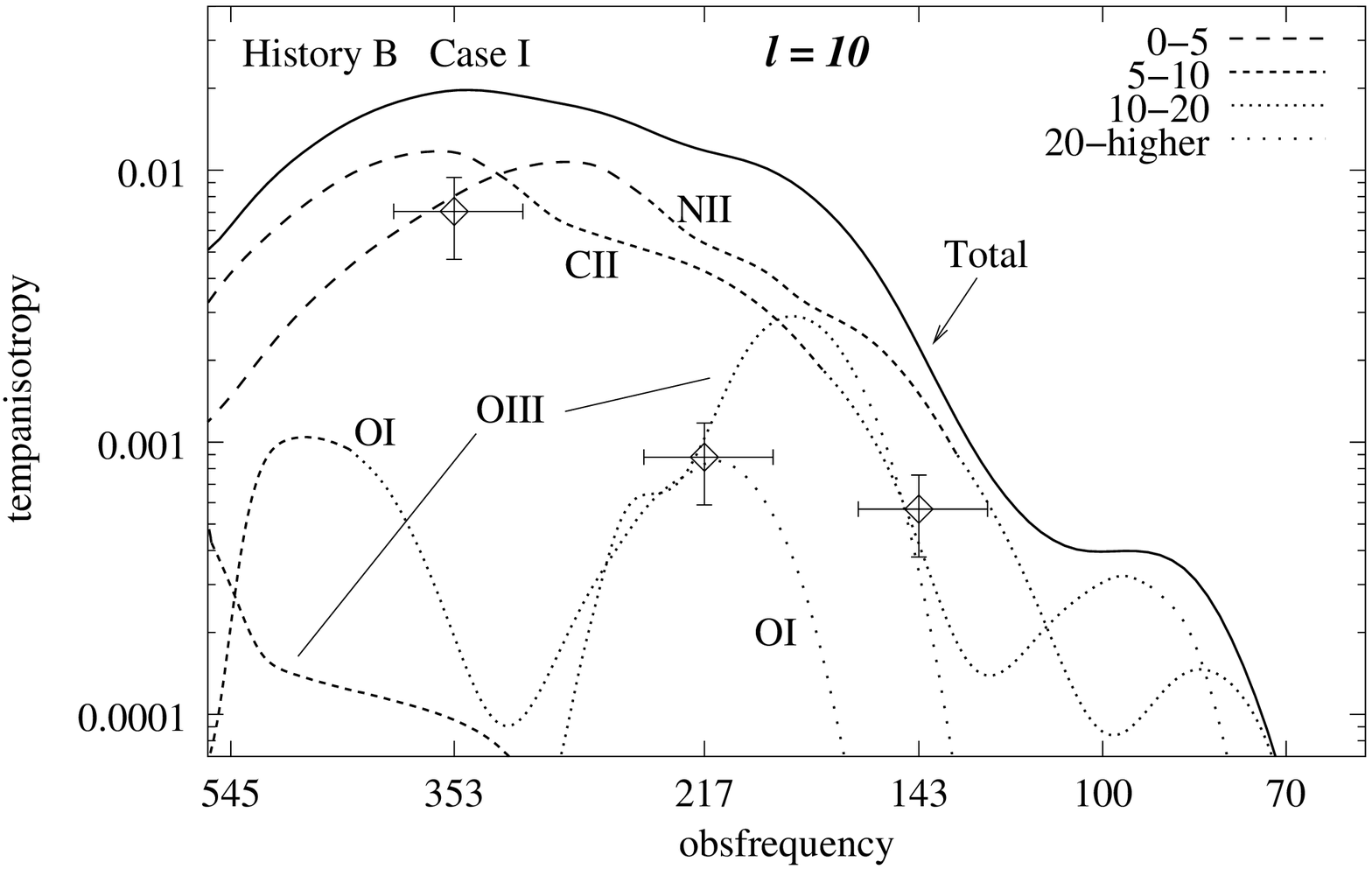}%
\includegraphics[width=9cm, height=6.5cm]{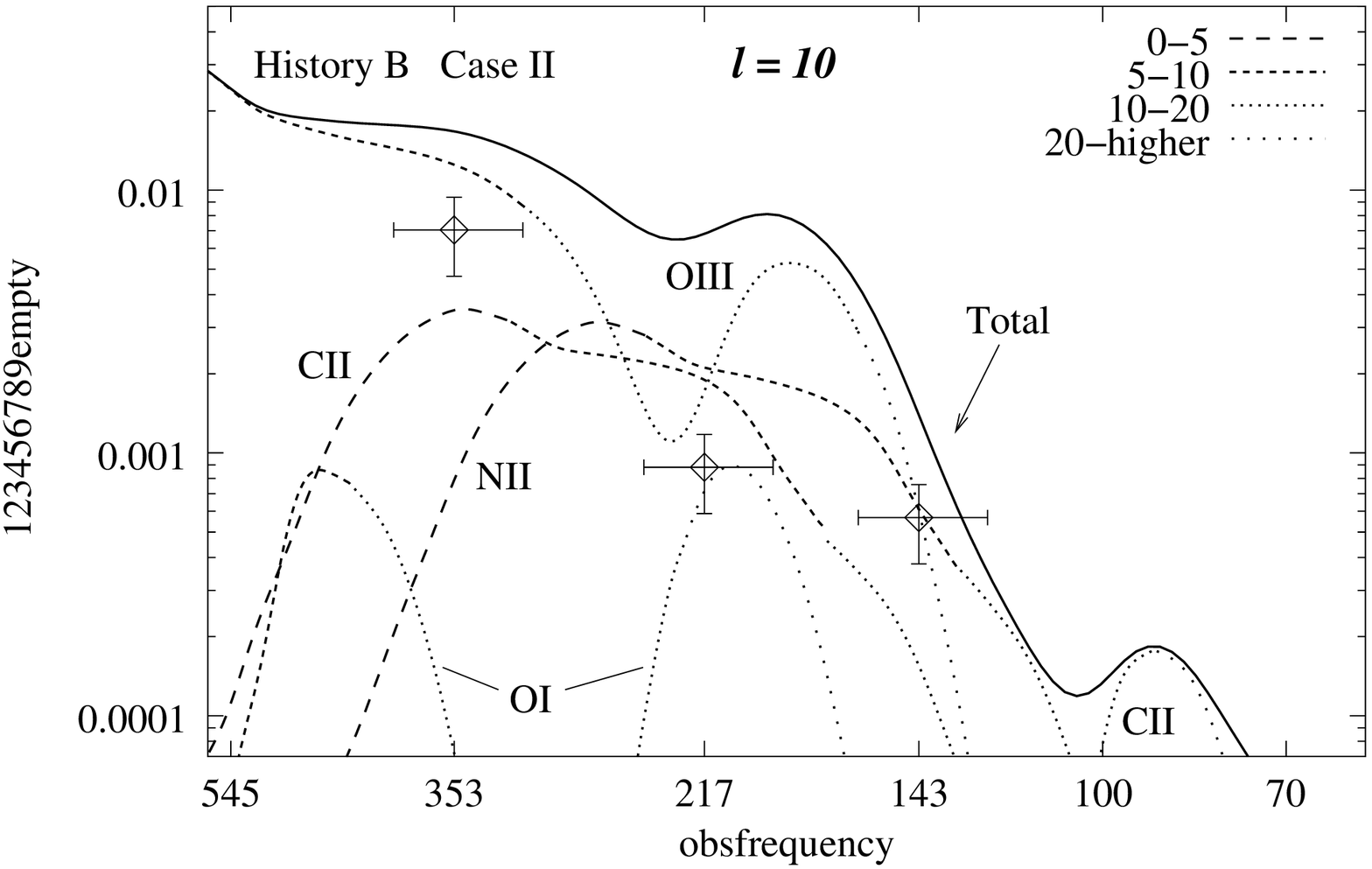}

\caption {
Frequency dependence of temperature anisotropy generated by scattering 
from fine-structure lines.  
We take two ionization histories and the high abundance case from
figure (\ref{skplot}),  
and show the contributions of different lines in different frequency
range, for a fixed multipole $l=10$. 
Four different
redshift ranges are marked for each species to emphasize the epochs
where the dominant contribution from each lines are coming.  
The sensitivity limits of Planck HFI channels are marked by the
crosses. 
The HFI limits have
been improved by a factor of $\sqrt{\Delta l}$ by averaging over in
the multipole range $l=7-16$ . The sensitivity limits are for
3 standard deviation detection, and the {\it y-errorbars} correspond to
$1\sigma$ error in $3\sigma$ detection (we recall that 
$\sigma = 3 \sqrt{ \sigma_{C_l(probe)}^2 + \sigma_{C_l(ref)}^2 }$,
where the reference channel is fixed at HFI 100GHz).   
The {\it x-errorbars}
corresponds to the wide bandwidth ($\sim$25\%) of Planck HFI channels. }

\label{mainplots}
\end{figure*}

In Fig.3b and 3c , we sketch
the relative ion fraction of the three most important atomic and ionic
species under above-mentioned ionization history. We show two
different reionization scenario: one for relatively 
cold stars when production OIII is less efficient ({\it case I}), 
and the other for hot stars and quasars 
which are able to keep oxygen fully ionized at all intermediate
redshifts ({\it case II}).  
Line of OI should give us an information about creation of oxygen
before Universe was strongly ionised. Relative growth of CII line
(see Fig.2b) will mark the time when carbon will be ionised in large
ionised regions which do not completely overlap or will be partially
ionised everywhere (we can not dinstinguish this two variants of
ionization history using large angle observations).  
Potential of CI ionization (I=11.26 eV) is lower than 
that of HI. Therefore CII fraction might be higher than that of HII
and OII in the beginning of secondary ionization. At the same time OI
fraction should follow that of HI because ionization potentials are so
close (I=13.62 and 13.60 eV correspondently).
Pop III stars should be very hot, and therefore they 
are able to ionize helium
early enough. Simultaneously OIII should become abundant ion 
because ionization potential of OII (I= 35.12 eV) is higher 
than that of HeI but smaller then that of HeII.

\begin{figure}
\psfrag{tempanisotropy}{$l(l+1)/2\pi \ |\delta C_l|$ \ \ ($\mu$K$^2$)}
\psfrag{multipole}{multipole, $l$}
\centering

\includegraphics[width=8.5cm, height=6cm]{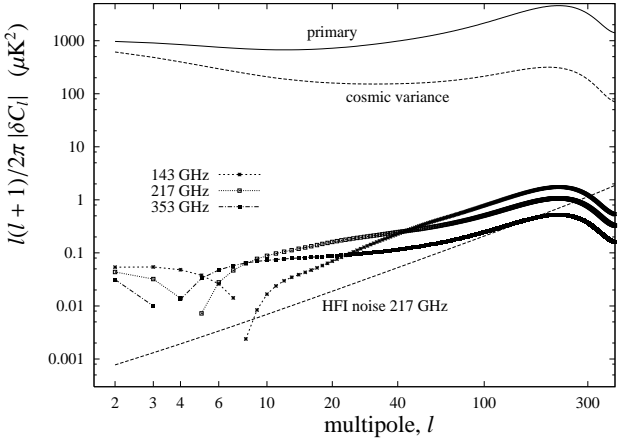}
\caption {Angular dependence of temperature anisotropy 
generated by scattering 
from CII 158$\mu$ fine-structure line. Shown here are three cases 
of $\left| \delta C_l \right|$ as
should be observed from Planck HFI 143 GHz (scattering at $z=12$), 217
GHz (scattering at $z=8$) and 353 GHz (scattering at $z=4$) channels,
(using 100 GHz channel as reference)for 10\% abundance of CII ions 
(with respect to solar) at all three redshifts. 
Also shown are the
primary anisotropy, $C_l$, the cosmic variance limit, and noise level
for 217 GHz channel (with respect to 100 GHz) for comparison. }

\label{myplot}
\end{figure}

\begin{figure*}
\psfrag{temperatureanisotropy}{$l(l+1)/2\pi \ |\delta C_l|$ \ \ ($\mu$K$^2$)}
\psfrag{obsfrequency}{$\nu_{obs}$ \ \ (GHz)}
\centering
\includegraphics[width=8cm, height=5cm]{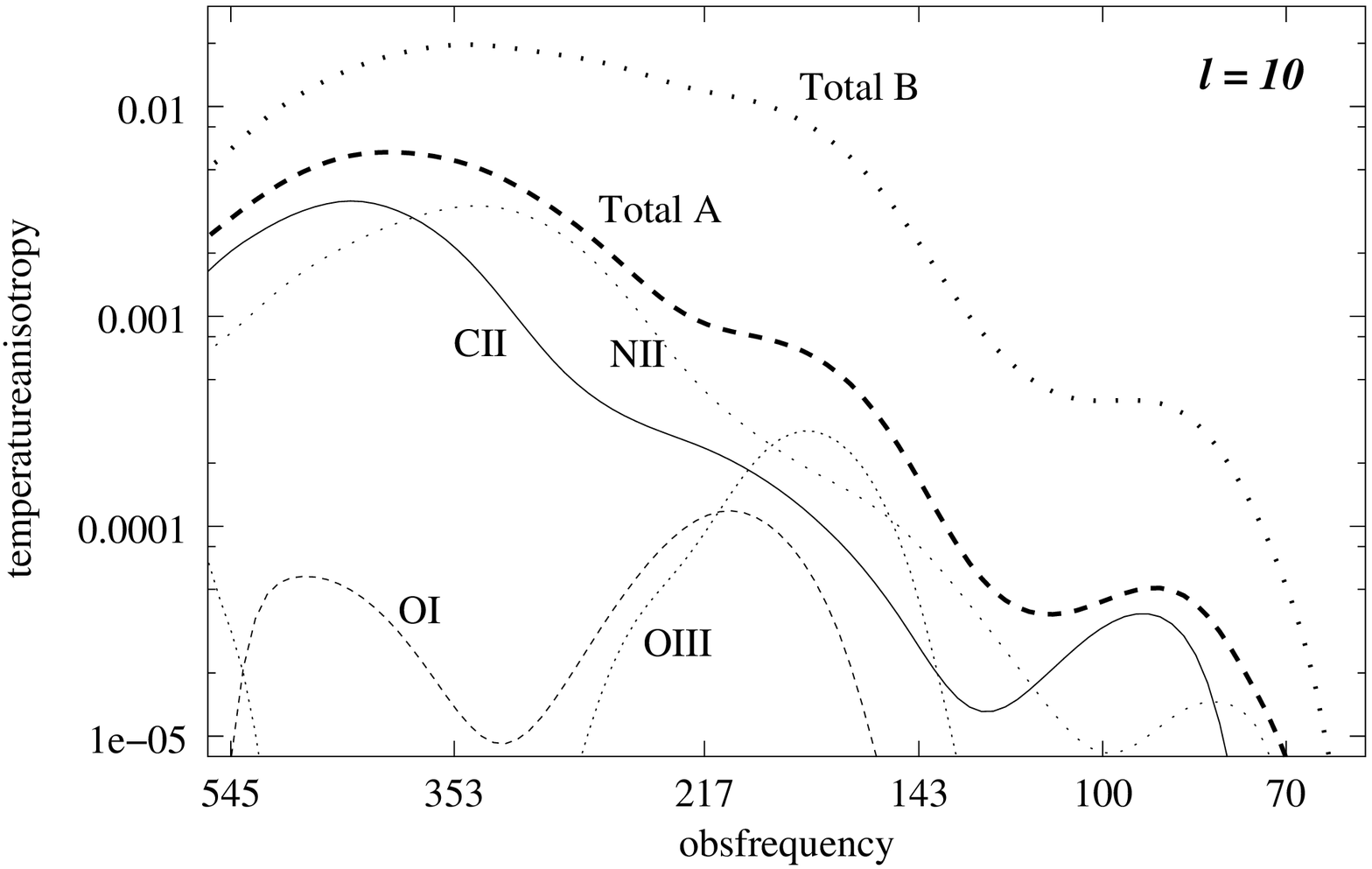}%
\hspace*{.3cm}
\includegraphics[width=8cm, height=5cm]{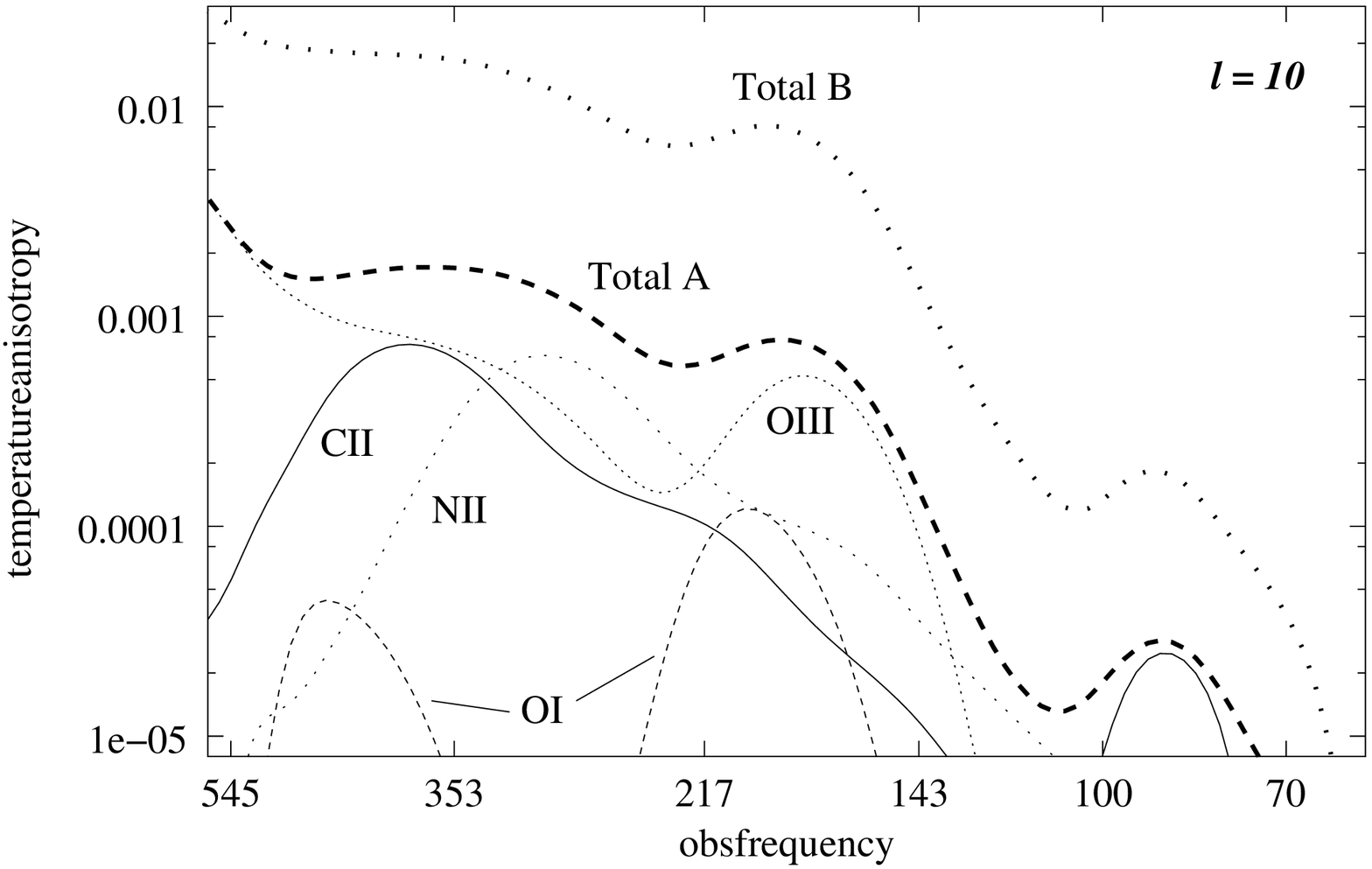}

\vspace*{.5cm}

\includegraphics[width=8cm, height=5cm]{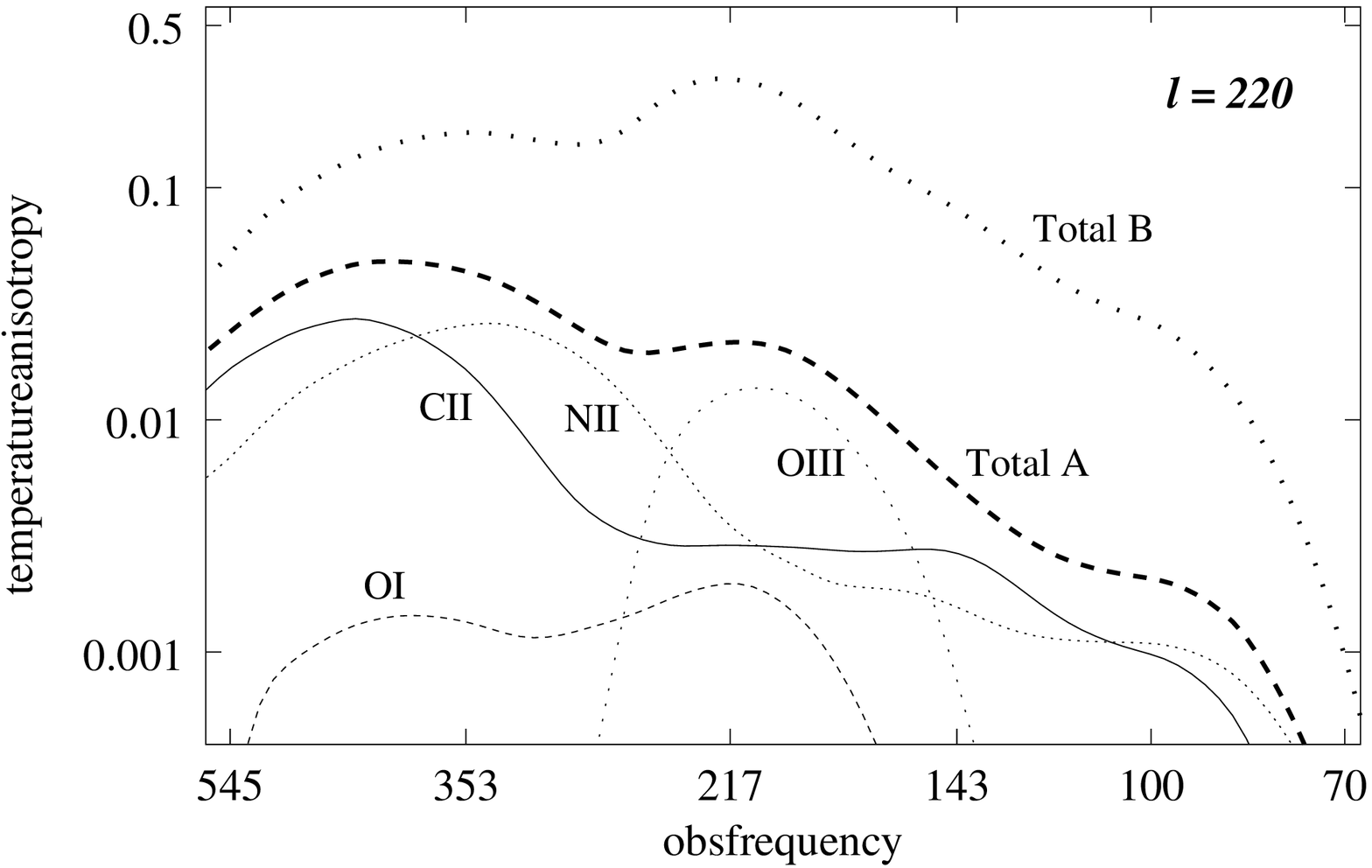}%
\hspace*{.5cm}
\includegraphics[width=8cm, height=5cm]{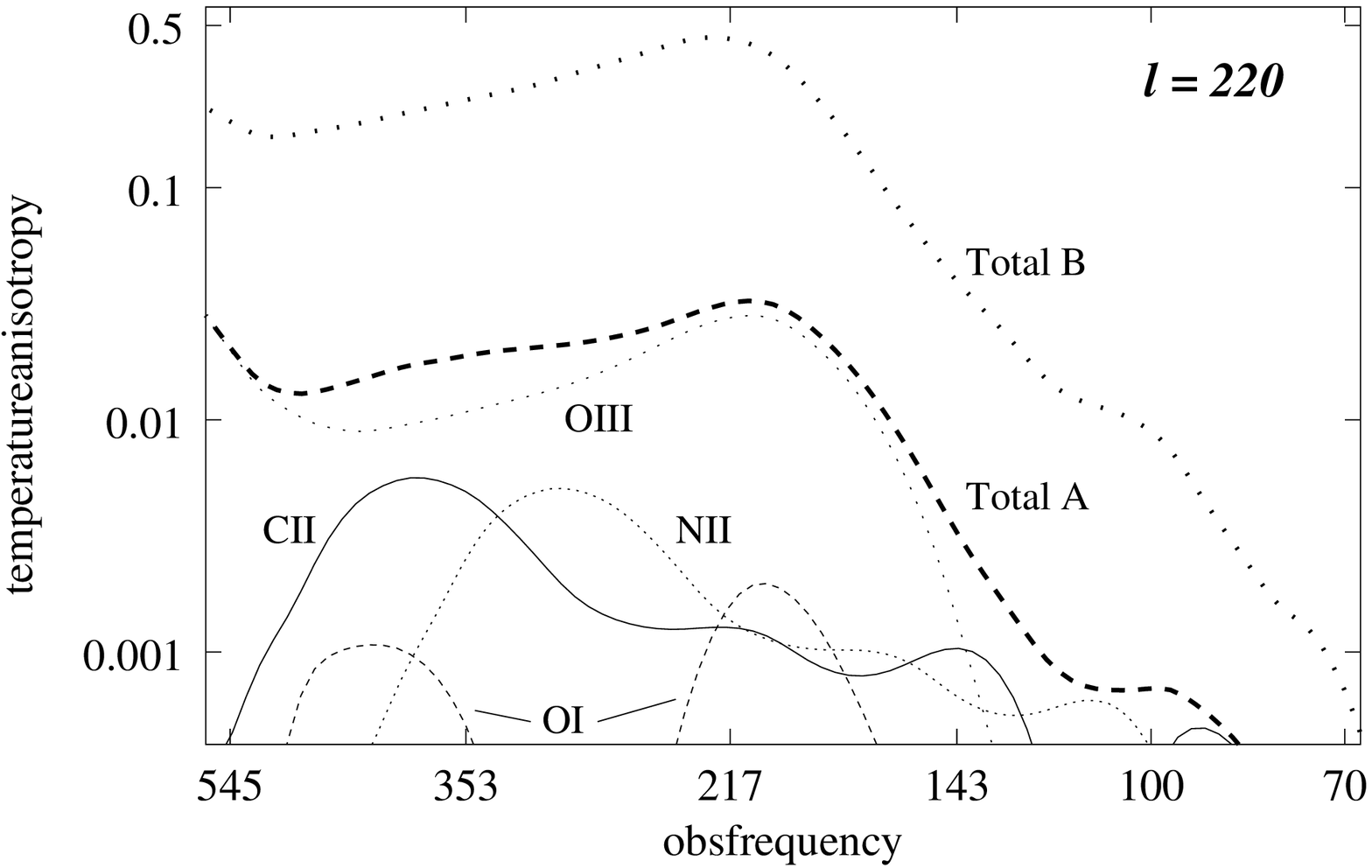}

\vspace*{.5cm}

\includegraphics[width=8cm, height=5cm]{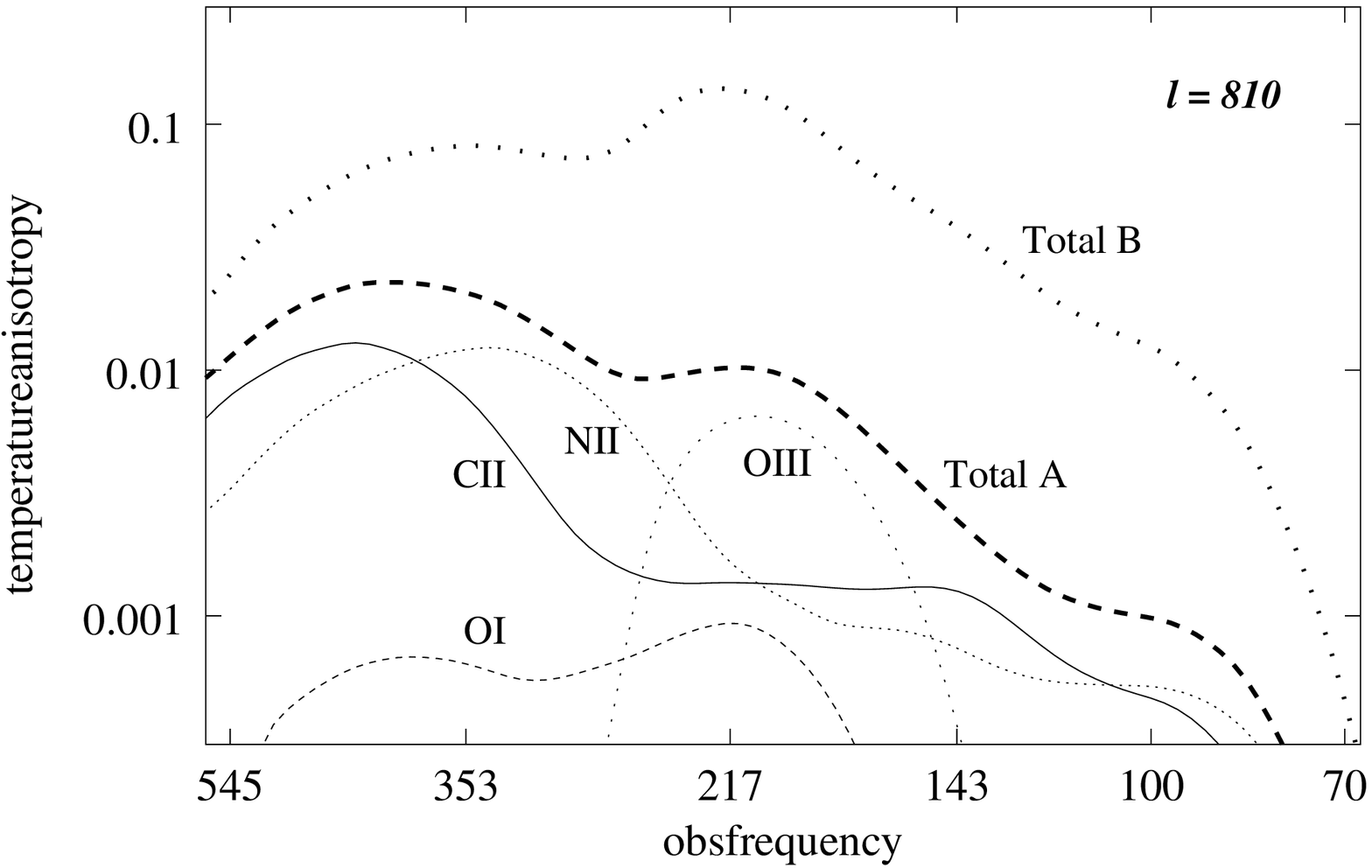}%
\hspace*{.5cm}
\includegraphics[width=8cm, height=5cm]{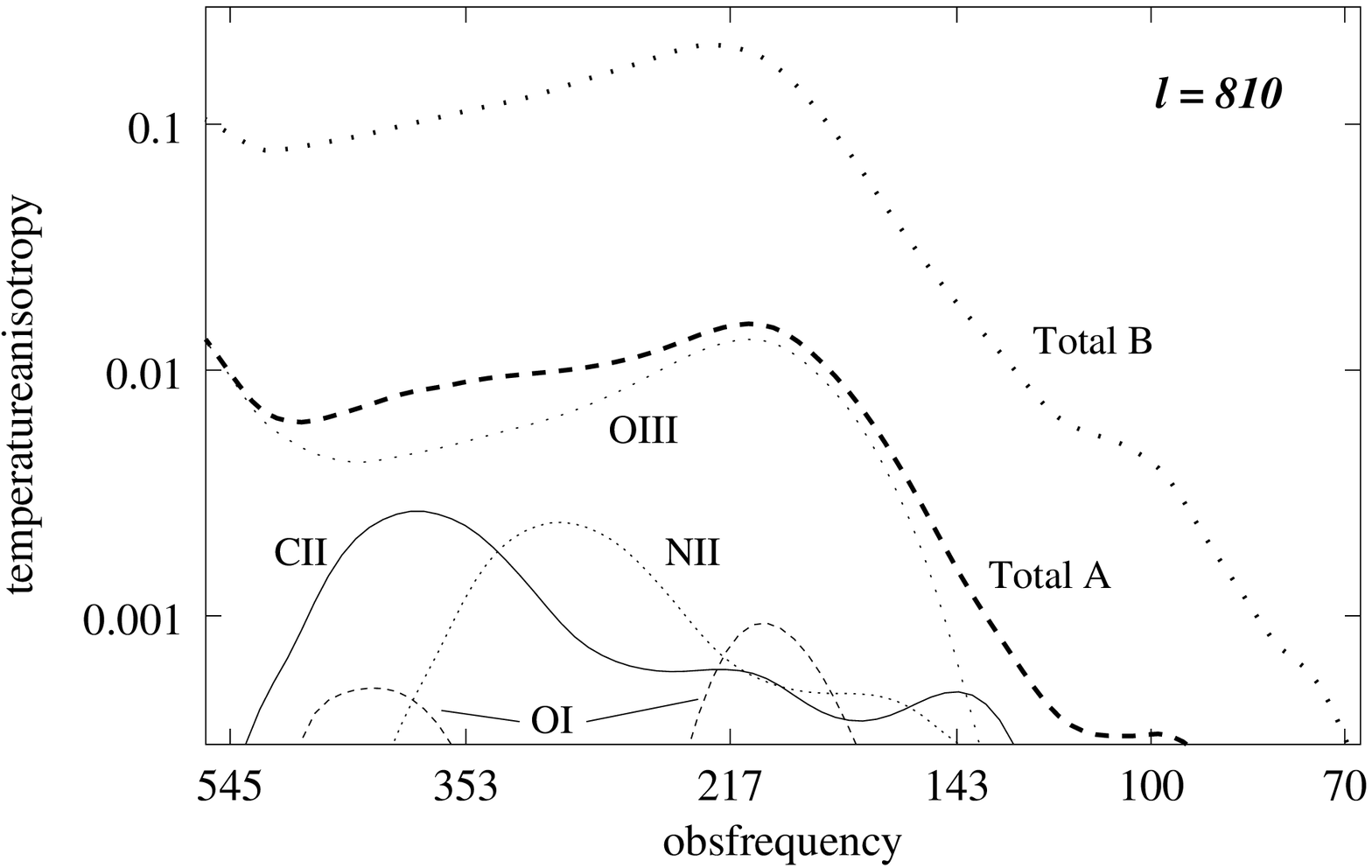}

\caption{Temperature anisotropy generated by atomic and ionic fine
structure lines at different frequencies in the case of three 
angular scales of observation.   
The figures on {\it left} are for Case I of Fig.3,  
which is for reionization by stars unable to ionize OII. Figures on
{\it right} are for Case II, which is for 
reionization by hot stars and quasars and
hence have high OIII fraction.  Each figure gives the total
contribution for both history A and history B. However, the separate
contributions from different lines are given for history A only. We
present results for three angular scales: $l=10$ ({\it top}), $l=220$
(first Doppler peak, in {\it middle}), and $l=810$ (third Doppler
peak, at {\it bottom}). The scattering-generated $\delta C_l$-s are
proportional to the intrinsic $C_l$-s 
for higher multipoles, but 
magnitude is slightly less at the position of third peak 
because of the reduced power in primary spectrum from $l=220$ to
$l=810$. }


\label{panels}
\end{figure*}

\begin{figure*}
\psfrag{temperatureanisotropy}{$l(l+1)/2\pi \ |\delta C_l|$ \ \ ($\mu$K$^2$)}
\psfrag{obsfrequency}{$\nu_{obs}$ \ \ (GHz)}
\centering

\includegraphics[width=7.5cm, height=6cm]{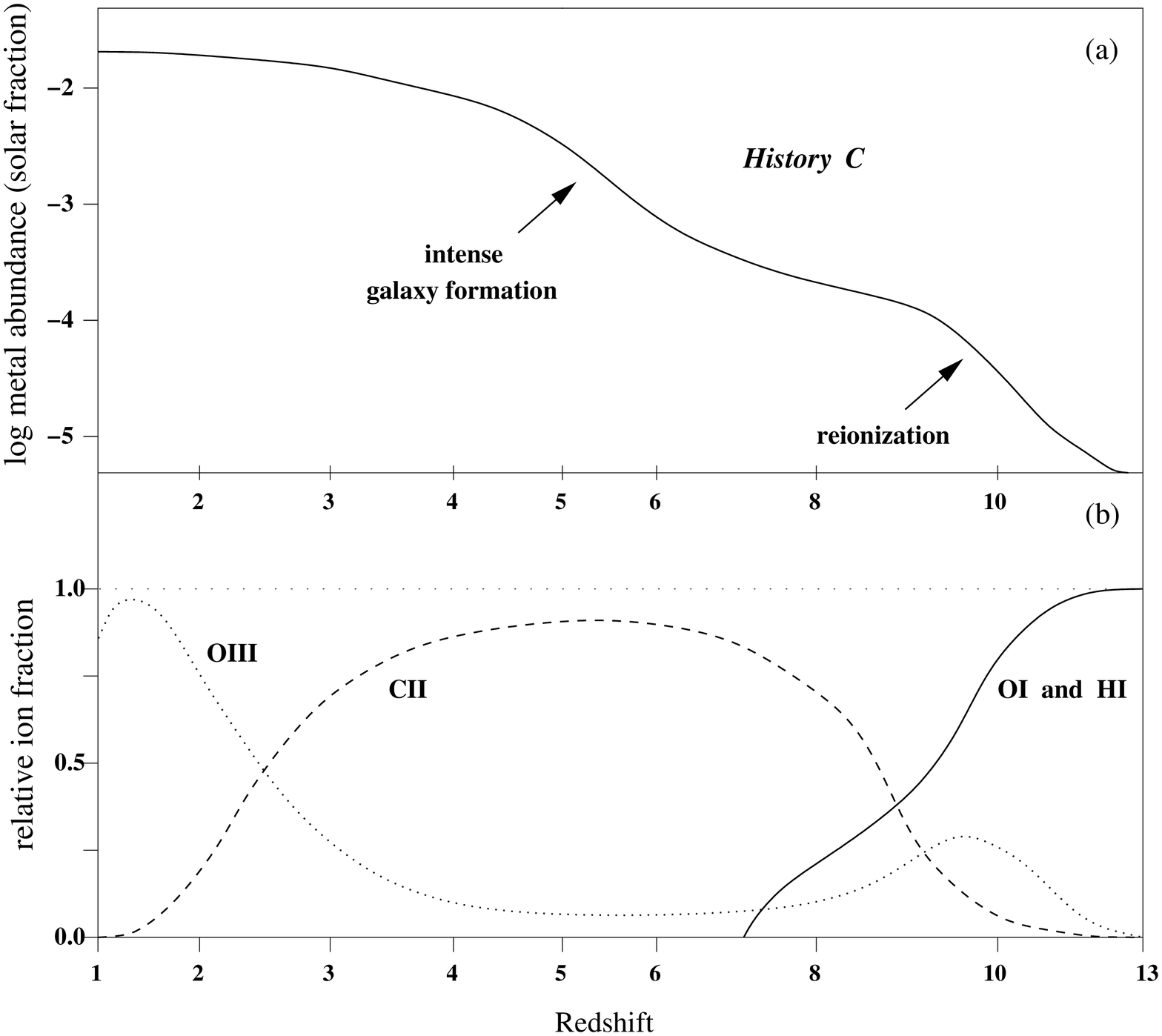}%
\hspace*{.4cm}
\includegraphics[width=8.1cm, height=6.1cm]{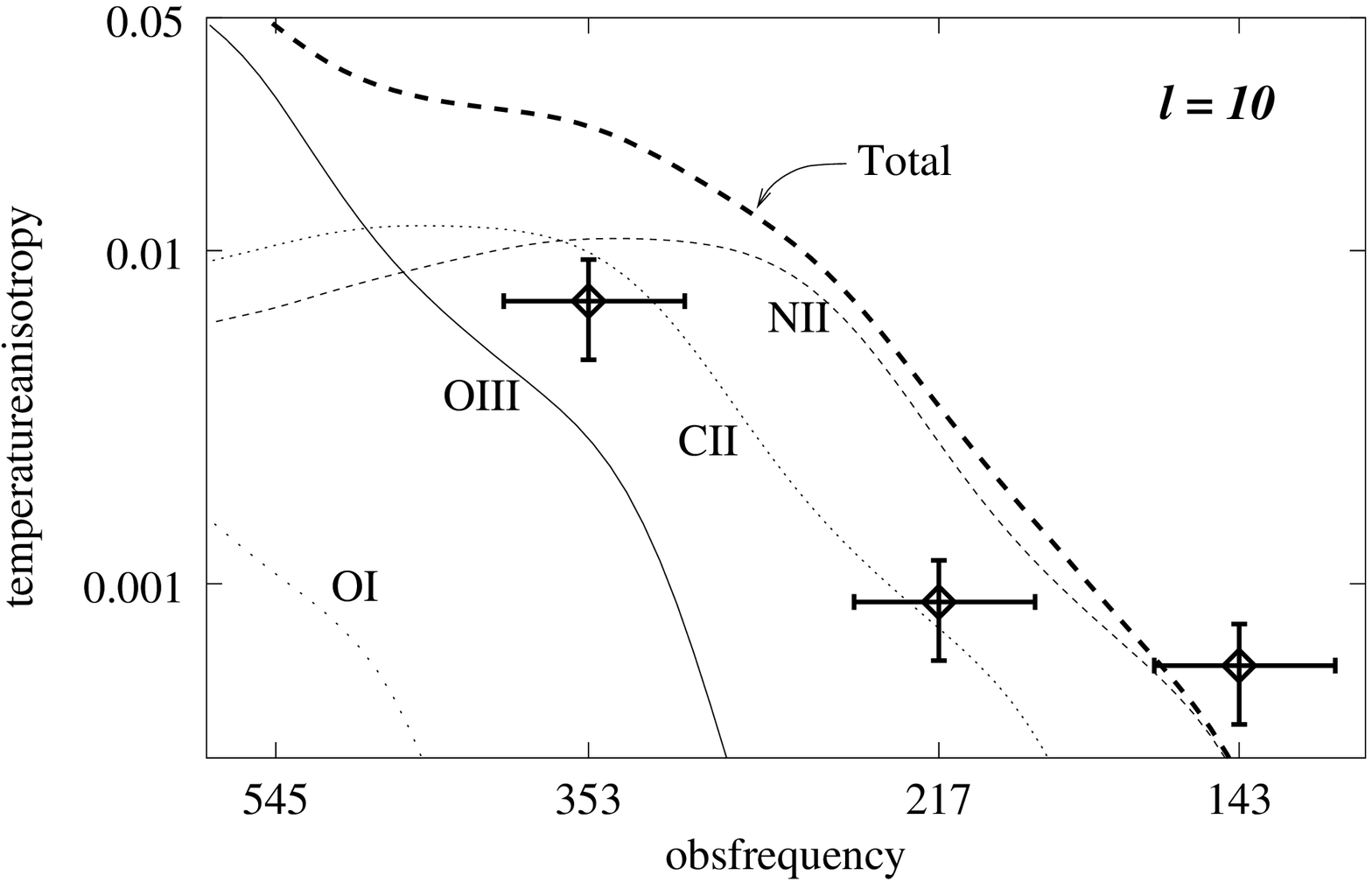}
\vspace*{.7cm}

\includegraphics[width=8.2cm, height=5cm]{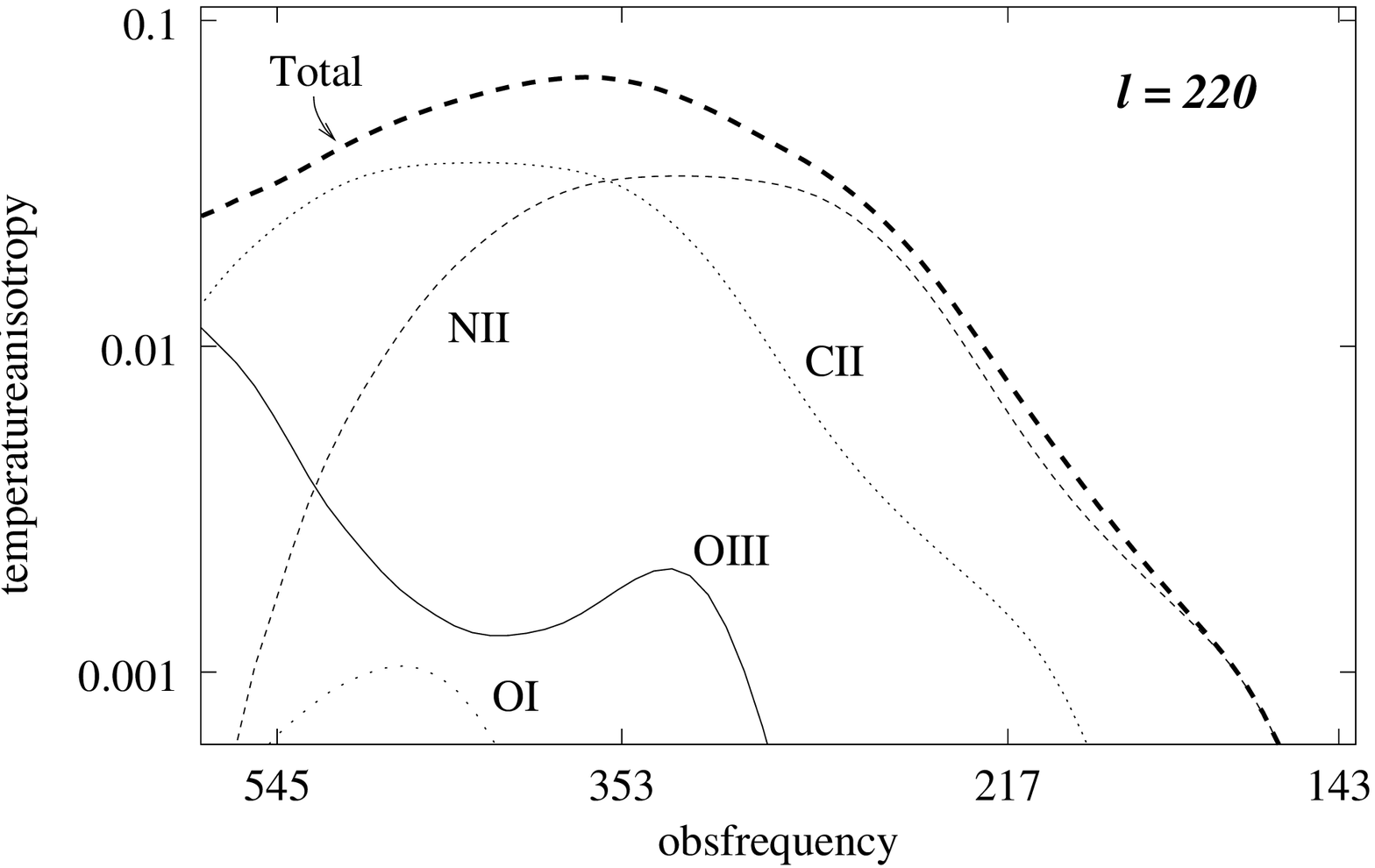}%
\hspace*{.4cm}
\includegraphics[width=8.2cm, height=5cm]{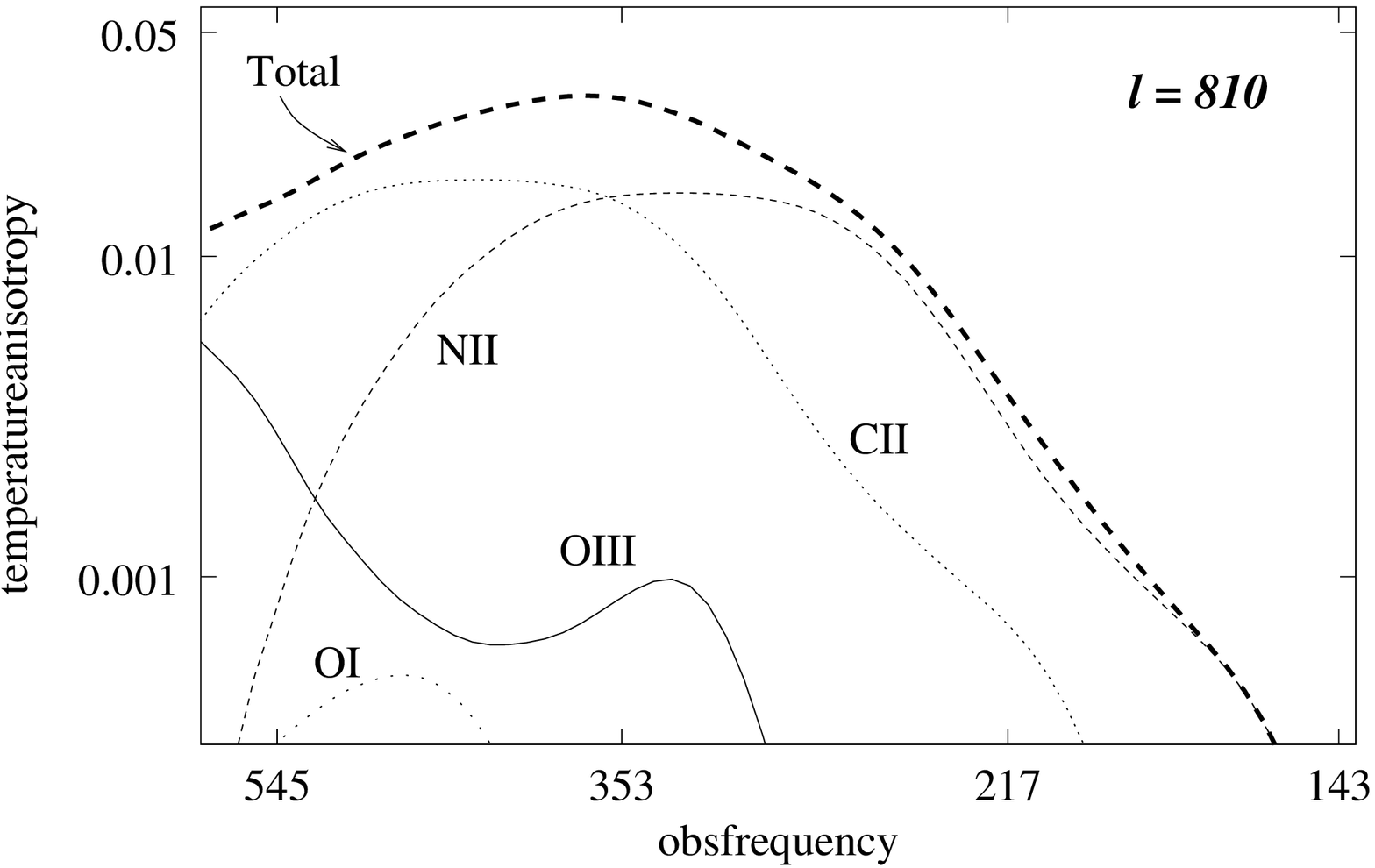}

\caption{Late ionization and enrichment history of the universe. 
{\it Top, left:} Sketch for the abundance history with late
reionization, 
where first metals are produced around $z \sim 10-12$. 
{\it Top, right:}
Temperature anisotropy for this abundance history generated at $l=10$.
Also shown are the Planck HFI sensitivity, where the noise levels have
again been improved by $\sqrt{\Delta l}$ in the multipole range
$10-20$. {\it Bottom left and right:} Temperature anisotropy for
$l=220$ and $l=810$, respectively.}

\label{sloanplot}
\end{figure*}

Fig.\ref{mainplots} shows the amplitude of the
predicted signal as a function of observing frequency, for a fixed
angular scale $l=10$. The contributions from four most important
lines, viz. CII 158$\mu$, NII $205\mu$, OI 63$\mu$ and OIII 88$\mu$,
are shown, along with their sum. 
We also show four different redshift ranges for each line, which
highlights the fact that contributions from CII and NII are higher
because their signal is coming from lower redshifts where abundance is
higher according to our chosen abundance history. 
In Fig.\ref{myplot} we show the angular dependence of the  
temperature anisotropy generated by resonant line scattering   
for the three important HFI channels. The abundance of CII
ion is kept fixed at 10\% solar, so that the lower observing frequency
(i.e. higher scattering redshift) has a higher value of optical depth
in the same line. The left-side of each curve is dominated by Doppler
generation of new anisotropy, and hence are positive. The righ-side is
dominated by suppression of primordial anisotropy and hence $\delta
C_l$-s are negative, the discontinuity in each curve shows the
interval where $\delta C_l$ changes sign. We see at low multipoles
both generation and suppression term tend to cancel each other. 
Especially due to the adoption of WMAP reionization model with high
optical depth ($\tau_{reio}=0.17$) in our computations, lines which
scatter CMB photons at redshifts $z \lesssim 18$ encounter a high
value of visibility function due to reionization, and therefore causes
a strong Doppler generation (but negative) in the multipole range $l =
20-100$. This is the cause for higher amplitude of the signal at 143
GHz. But at high multipoles $l>100$ the contribution from Doppler
generation is negligible, and the signal is simply proportional to the
primordial $C_l$-s, and the line optical depth, provided $\tau \ll
1$. Hence using data from table(\ref{hfitable}), 
and knowing the amplitude of primordial CMB
signal, one can immediately predict the amplitude of the effect at
small angular scales using formula(\ref{simple}).

Fig.\ref{panels} summarizes our results for three different angular
scales. As explained above, at low multipoles the effect is not
proportional to $C_l$, as can be seen from the two panels at
top. These low multipole results are important for future satellite
missions like Planck and CMBPol. In the middle, results for $l=220$
are shown, where effect is already proportional to $C_l$. The
amplitude of the signal is highest here due to the large amplitude for
the $C_l$ at the first Doppler peak. At the bottom we give results for
the third Doppler peak, or $l=810$, where signal drops by a factor of
$2$ from that of $l=220$. However, these angular scales ($\theta
\sim 13'$) are particularly suitable for future balloon and 
ground-based experiments with multi-channel narrow-band recievers.

Finally, in Fig.\ref{sloanplot}, we present an alternative
ionization and enrichment history of the universe, when there were no
production of heavy elements before $z \sim 12$. This is very
interesting because, even for such late reionization, if there is
moderate enrichment of the IGM around redshift $4-5$, we have the
possibility to detect our signal with Planck HFI around $l=10$. The
contributions from high energy oxygen lines are reduced because of the
absence of metals at high redshift, but long-wavelength lines of CII
and NII still can generate strong signal from low redshifts ($z
\lesssim 8$). The HFI sensitivities in this figure have again been
improved by averaging the noise in a multipole range of $\Delta l=10$
around $l=10$. As discussed before, the signal for $l=220$ and $l=810$
are proportional to the optical depths in these lines and the
primordial $C_l$-s.

Due to relatively small scattering cross-sections of the 
fine-structure transitions
under discussion, such
observations are sensitive to significant abundances of the atoms and ions
(for example when given species contributes from 10\% up to 100\% of
the corresponding element abundance), whereas the Gunn-Peterson effect
gives optical depth of the order of unity already when abundance of
neutral hydrogen is of the order of $10^{-5} - 10^{-6}$. Hence even if
the universe is completely opaque to Ly-$\alpha$ photons at redshifts 
$z \gtrsim 6$, ionic fine-structure lines like CII $157.7
\mu$ can probe the very first stages of patchy reionization process.
Detection of all three broad spectral features connected with strongly
redshifted OI, CII and OIII microwave lines
will open the possibility to trace the 
complete reionization history of the Universe.
In addition OIII line will proof the existence of 
extremely hot stars at that
epoch. 

The published level of noise of the Planck HFI shows that the three low
frequency channels of HFI are almost an order of magnitude more
sensitive than signal coming from cases I \& II of history B. Even in
the case of late reionization history (case C, Fig.\ref{sloanplot} ), 
Planck is about
4-times more sensitive than predicted signal. However, 
to be able to find contribution from at least three lines
simultaneously we need higher amount of frequency channels than
Planck HFI will have. 
A higher amount of frequency channels would certainly be 
possible for the next generation experiments like CMBPol even if
they are based on already existing technology (Church 2002).

There is the possibility 
that balloon and ground-based experiments will be able to
check the level of enrichment of the universe by heavy elements even
before Planck, observing at $l=220$ and $l=810$, for instance. At
these high multipoles, effect is directly proportional to the optical
depths in lines and the primordial CMB anisotropy.  
Hence using the data from table \ref{hfitable} and
Fig.\ref{skplot} (bottom), and 
the simple analytic relation $\delta C_l  \simeq - 2 \ \tau_{X_i} \
C_l^{primary}$, 
one can immediately give the effect in a first approximation. The high
signal-to-noise ratio of the primordial $C_l$-s around the first three
Doppler peaks will correspondingly give rise to strong signals in
scattering, and might become accessible 
through the tremendous sensitivity promised from 
the forthcoming balloon and ground-based telescopes like Boomerang,
ACT, APEX and SPT. 

CBI, VSA and BIMA interferometrs are studying successfully CNB 
angular fluctuations at low frequencies 30 GHz, there are plans to
continue observations on 45, 70 GHz, 90, 100 and 150 GHZ. 
CI lines with lambda 609 and 370 micron coming from high
redshifts 10-15 might contribute to the observed signal at 
low frequencies.

Beginning of the end of 60-ties theorists are discussing early star
formation due to isothermal perturbations, making possible production
of heavy elements at redshifts well above 100. This makes
interesting the lines with much shorter wavelengths, like $12.8 \mu$ 
Ne II line and many others, which might be contributing to the Planck
HFI spectral bands even from redshifts $\sim 100$.


\section{Effect of Foregrounds}

Before WMAP,
the amplitude and spatial and frequency scaling of foregrounds
had not been firmly established, and its
modeling was merely in a preliminary phase. After their first mission year,
WMAP's team have come up with a foreground model which is claimed to 
reproduce with a few percent 
accuracy the observed foreground emission 
(Bennett et al. 2003b). This and other future studies of foregrounds
may provide a characterization of these contaminants such that 
their effect on our method can be minimized.
We next proceed to estimate the impact of these contaminants on our method
by the use of current foregrounds models.

\begin{figure*}
\centering
\hspace*{-.5cm}
\includegraphics[width=14cm, height=8cm]{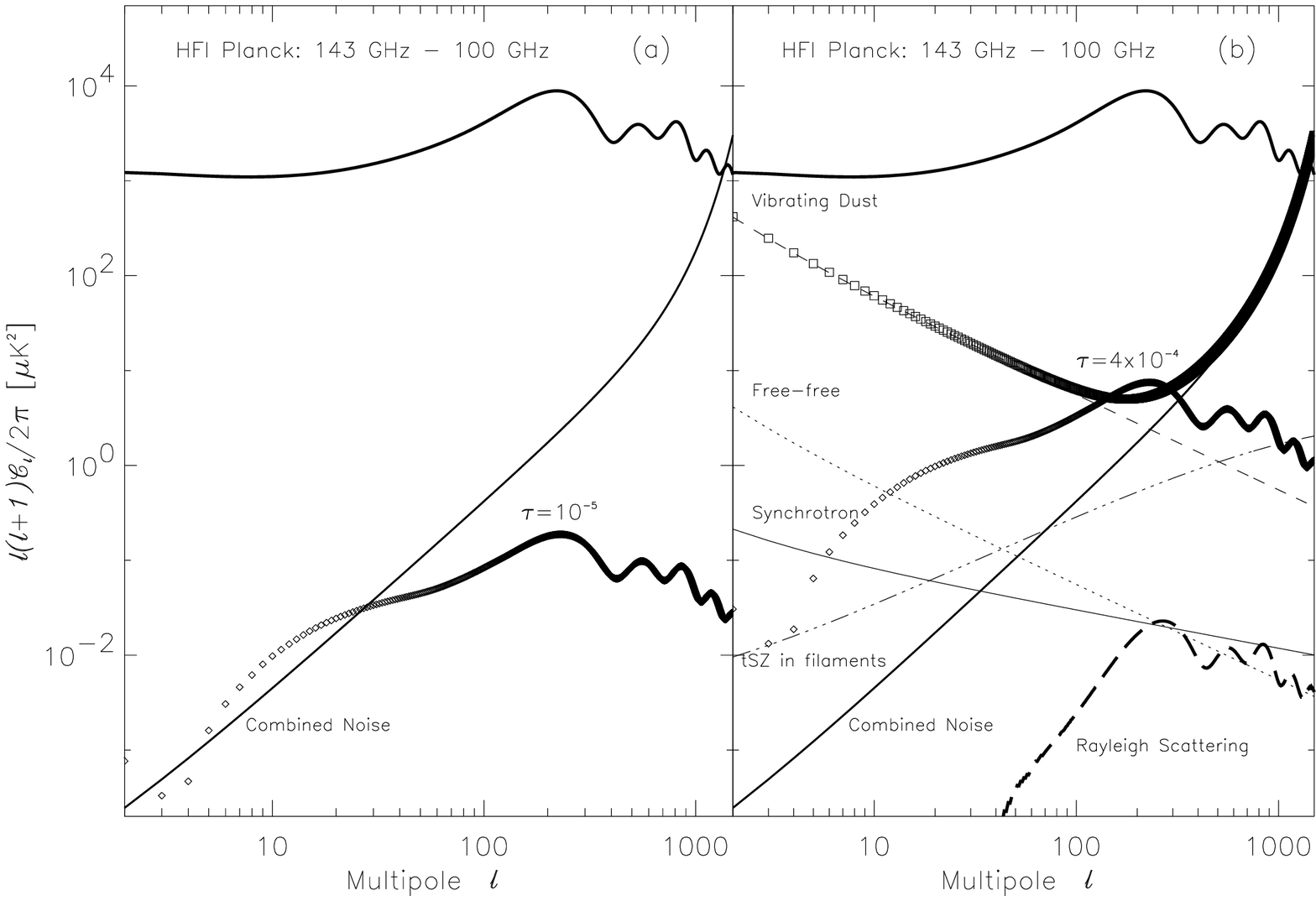}
\caption{	{\it (a)} 
Example of optical depth due to a resonant transition which can be
detected by the HFI 100 GHz and HFI 143 GHz PLANCK channels in the
absence of
foregrounds. The upper thick solid line gives the reference model CMB power
spectrum. Diamonds show the expected difference in the power spectrum from
 both channels due to the resonant scattering placed at $z=25$ with
 $\tau=1\times 10^{-5}$. 
{\it (b)} Presence of foreground contamination after subtracting HFI 100GHz
 channel
from HFI 143GHz channel. The reference $\Lambda$CDM CMB power spectrum
is shown in thick solid line, whereas the instrumental noise after the
map subtraction is shown in solid intermediate-thickness line. All thin
lines refer to the foreground model as quoted as in T00: vibrating dust
emission is shown by a dashed line, whereas free-free and synchrotron
are given
by a dotted and a solid thin line respectively. tSZ
effect associated to filaments and superclusters is given by the 
triple-dot-dashed line. Rotating dust gives negligible contribution at these
frequencies. Rayleigh scattering introduces some frequency-dependent 
variations
in the $C_l$'s, but well below the noise level (thick dashed line at the
bottom right corner).
Diamonds show the amplitude of the change in the CMB power
spectra induced by resonant species placed at $z=25$ and 
$\tau=3\times 10^{-4}$.}

\label{fig:foregrounds}
\end{figure*}

We have adopted the {\it middle-of-the-road} model
of Tegmark et al. 2000 (hereafter T00). This model 
studies separately the contribution coming from different
components, giving similar amplitudes for those which are also modelled by
the WMAP's team, (Bennett et al. 2003b).
In this model, we will consider the contribution of five foreground
sources, namely synchrotron radiation, free-free emission, dust emission,
tSZ effect (Sunyaev \& Zel'dovich 1972)
 associated to filaments and 
superclusters of galaxies, and Rayleigh scattering. 
 The $l$-dependence of the power spectra was
 approximated by a power law for all foreground components, (except for tSZ,
for which the model was slightly more sophisticated).
The frequency dependence observed the physical mechanism behind each
contaminant: simple power laws were adopted for free-free and synchrotron,
whereas a modified black body spectrum was used for dust. For tSZ, the 
frequency dependence of temperature anisotropies is well known in the
non-relativistic regime. All details about this modeling can be found at
T00. We are neglecting the contribution from the SZ effect generated in
 resolved clusters of galaxies which can be removed from the map. 
We are also assuming that all resolved point sources 
are excised from the map, and that the contribution from the remaining 
unresolved point sources ($\sigma_{ps}$ in eq.(10) of T00) can be lowered 
down to roughly the noise level.
This may require the presence of an external point-source
catalogue.

Fig.\ref{fig:foregrounds}a shows the expected precision level when
foreground contaminations can be neglected, and 
Fig.\ref{fig:foregrounds}b shows the effect of foregrounds in our
differential method to detect the presence of resonant species. Together
with the power spectrum of our Standard $\Lambda$CDM model, (thick solid
line at the top), we show the contribution from residuals of all 
foregrounds components under consideration, obtained after subtracting
 the HFI 100 GHz channel power spectra from the HFI 143 GHz one.
The thin solid line corresponds to synchrotron emission,
free-free emission is given by the dotted line; the dashed line gives
the contribution of dust through vibrational transitions. The tSZ associated
to filaments is shown by the triple-dot-dashed line, and the contribution
from the combined noise of both channels is shown by the solid line of
intermediate thickness which crosses the plot from the bottom left to 
(almost)
the top right corners. At these frequencies, the contribution of rotating 
dust is negligible, and the most limiting foreground is dust emission 
by means
of vibrational transitions. Finally, the last frequency dependent
 contribution
that we have considered corresponds to Rayleigh scattering. As shown in
Yu, Spergel \& Ostriker 2001, hereafter YSO, Rayleigh scattering of CMB 
photons with neutral Hydrogen atoms
introduce frequency dependent temperature fluctuations. However, 
provided that
$\tau_{Rayleigh} \propto \nu^4$, this process
is only effective at high frequencies ($\nu \gtrsim 300$GHz), causing 
deviations
of measured power spectrum from the model $C_l$'s of the order of a few
percent. YSO characterized the Rayleigh scattering by introducing in the
CMBFAST code a frequency dependent effective optical depth 
and a frequency integrated drag force exerted on the baryons. This last
modification, which was neglected when considering resonant scattering,
 couples the evolution of the different temperature multipoles
($\Delta_{T,l}$) for different frequencies. However, this coupling is
exclusively due to the dipole term $\Delta_{T,1}$, which, at the light of 
their results, is very similar for all frequencies. Taking
the same dipole term for every frequency allows us evolving a 
separate system
of differential equation for each frequency. The frequency dependent changes
in the angular power spectrum, ($\delta C_l / C_l (\nu )$), 
 obtained under this approximation are
almost identical to those in YSO. For each channel pair under 
consideration, we computed the residual in the power spectrum
difference due to these frequency dependence temperature 
anisotropies. We checked that this effect was always subdominant. In
Fig.\ref{fig:foregrounds}b, the impact of Rayleigh scattering is shown
by a middle thick dashed line at the bottom-right corner of the panel,
 for the
particular comparison of the 100GHz and 143GHz channels.

 The addition of all contaminants (foregrounds + 
noise) is given by squares. Diamonds show the change in the power spectrum 
associated to a resonant transition placed at $z=25$ for 
optical depth $\tau=3\times 10^{-4}$. 
After comparing with Fig.\ref{fig:foregrounds}a we can appreciate how
the presence of foregrounds affects our method in two different ways: 
{\it a)} it increases the minimum $\tau$ to which the method is sensitive
by about two orders of magnitude, {\it b)} it changes the range of
multipoles to look at, due mainly to the large typical angular size of 
dust clouds. In the particular case of HFI 100GHz-143GHz channels,
one should focus on the range $l\in [100,200]$ once the dust component is
included rather than in $l\in [3,10]$ for the foreground-free case. 
Recurring again to the linear dependence of $\tau$ versus the abundance,
eq.(\ref{sobolev}), we see that this model of 
foregrounds increases by about two orders of magnitude the minimum 
abundance 
on which future CMB missions will be able to put constraints.



\section{Conclusions}

In this paper, we
 have computed the effect of heavy elements on the
Cosmic Microwave Background. We have selected those species
which might leave a greater signature
in the CMB power spectrum, and analyzed the possibility of constraining
their abundances by the use of forthcoming CMB data. 

Under the approximation of negligible drag force, we have computed what 
are the changes in the CMB angular power spectra introduced by resonant
species placed at redshifts which are dependent on the observing 
and resonant frequencies. The overall effect can be decomposed into a
damping or suppression of the original CMB temperature fluctuations, and
a generation of new anisotropies. For the optically thin limit to which we
have restricted ourselves, damping dominates over generation of new 
anisotropies in the intermediate and high multipole range ($l \gtrsim 20$). 

These changes in the CMB are of very small amplitude ($0.01-0.3\mu K$), and
could be distinguished from the CMB component by means of their frequency 
dependence. Indeed, a comparison with respect to a 
{\it reference} channel containing only non-frequency
dependent temperature fluctuations could be used to quantify the amount
of angular power introduced by the resonant species. However, this would only
be feasible if either the other frequency-dependent components are negligible
or characterized with extreme accuracy. The possible presence of foregrounds,
galactic of extragalactic, whose amplitude and spectral behavior is still
under characterization, is a serious aspect to be taken into account.
 Other technical challenges, such as the calibration of 
different channels and the enormous sensitivity required, should be accessible
from the next generation of detector technology.

We have obtained limits for abundances of heavy elements 
when complete foreground removal is possible, but also 
have shown values when 
all foregrounds are present in the sky map.  
Our analyses 
have particularly focused on PLANCK HFI channels, whose very low noise
levels give extremely strong limits on abundances.  
It is easiest to observe the proposed effect with HFI at angular
scales of $\theta \sim 20^{\circ}$ or $l \sim 10$ because of the very
low noise of the first three HFI channels at these multipoles.
Using the 143 \& 217 GHz
channels of HFI, with the 100 GHz channel as reference, limits as low
as $10^{-3}-10^{-4}$ solar abundance were obtained for 
atoms and ions of the most important elements like carbon, nitrogen
and oxygen in the redshift range $[5,30]$. At higher multipoles
($l > 200$), we have shown that future balloon and ground-based 
experiments like ACT, APEX and SPT can put similar constraint on
abundances, where the predicted signal is stronger because of the higher
amplitude of the primordial CMB signal, and effect will be directly
proportinal to the optical depths in lines.  
The presence of foregrounds makes
all these limits about a factor of $10^2$ worse, but that must be
considered as the most pessimistic scenario.

In this paper we have considered scattering of CMB photons by low
density gas everywhere in the universe, where optical depths in 
the relevant atomic or molecular lines are very low and resonant 
scattering effect dominates over line emission. 
The effect connected with scattering from atomic and ionic fine-structure
transitions depends on the amount of of atoms/ions in the ground
state, which is determined by the excitation temperature. In Appendix
B we showed that even up to very high over-density of the plasma, the
change in excitation temperature from the CMB temperature is
negligible, and our simple approximation is valid. 
The clustering of heavy atoms and molecules in clouds should change the
predictions on the modifications of the CMB power spectrum
at smaller angular scales, and will be discussed in a subsequent work.

\medskip

\medskip
\begin{acknowledgements}
The authors thank M. Zaldarriaga and J. Chluba for useful comments, 
and especially L. Page for discussing the possibility to detect this
effect at high $l$-s with the tremendous sensitivity of forthcoming
{\it Atacama Cosmology Telescope}.  KB 
is grateful to Dr. D. Lutz of Max-Planck-Institute 
for Extraterrestrial Physics for giving access to the ISO line list.
CHM acknowledges the financial support provided through the European
Community's Human Potential Programme under contract HPRN-CT-2002-00124, 
CMBNET.

\end{acknowledgements}

\appendix
\section[]{}

\begin{flushleft}
\small {\it METHOD OF COMPUTATION:}
\end{flushleft}
 Our approach to model the coupling between these heavy species
and CMB photons will be similar to that outlined in Zaldarriaga 
\& Loeb (2002, ZL02).  
In their paper they discussed modifications of CMB
spectrum by presence of primordial Lithium atoms. Here we extend their
analysis to other atoms and ions. We also discuss
the necessary changes that are enforced while extending this method to
all resonant species.

As a first step, we describe the modifications introduced in the 
CMBFast code in order to compute the effect of resonant transitions.
We consider a given resonant transition {\it i} of a given species
{\it X}, with a resonant frequency $\nu_{X_i}$. For a fixed observing
frequency $\nu_{obs}$, the redshift at which that species interacts with the
CMB is $1+z=\nu_{X_i} / \nu_{obs}$, and its opacity can be 
written, in
general, as $\dot{\tau}_{X_i} = \tau_{X_i} {\cal P} (\eta )$, with
a normalized profile function $\int_0^{\eta_0} d\eta' {\cal P(\eta')}=1$. We
shall model this profile with a gaussian: 
\begin{equation}
\dot{\tau}_{X_i} = \tau_{X_i} \; 
			\frac{\exp{ \left(-\frac{\left( \eta - 
 \eta_{X_i} \right)^2} {2\sigma_{X_i}^2} \right) }}{\sqrt{ 2\pi
			\sigma_{X_i}^2}}
\label{gauss}
\end{equation}
where $\tau_{X_i}$ is the optical depth for the specific transition, 
$\eta_{X_i}$ is the conformal
time corresponding to the redshift $1+z_{X_i}=\nu_{X_i} / \nu_{obs}$ where
scattering takes place, and $\sigma_{X_i}$ is the width of the gaussian. 
For a fixed transition, this width
should be given by the thermal broadening of the line.
For the sake of simplicity, we shall take 
$\sigma_{X_i} \sim 0.01 \eta_{X_i} $.

Once the line optical depth has been characterized, (as shown in
Section 2), the new
opacity term with a gaussian profile is added to the standard Thompson 
opacity,

\begin{equation}
\dot \tau = a n_e \sigma_T  +  \sum_{i} \tau_{X_i} \frac{\exp{ \left( 
-\frac{\left( \eta - \eta_{X_i} \right)^2}{2\sigma_{\eta_{X_i}}^2}
\right) }}{\sqrt{
2\pi \sigma_{\eta_{X_i}}^2}},
\label{opacity}
\end{equation}

and this in turn is used to
compute the visibility function $\Upsilon (\eta )$, defined as 
\begin{equation}
\Upsilon(\eta) = {\dot \tau}(\eta) e^{-\tau (\eta)} .
\end{equation}

This scattering introduces a frequency dependent term in the evolution
equation for the photon distribution function, which results in a
frequency dependence drag-force. However, based on the same arguments of
ZL02 , we can safely ignore this drag-force as long as
the characteristic time of drag exerted by these species is far larger
than the Hubble time, which is indeed the case due to the low optical
depths under consideration. 
These are the modifications we  
introduced in the CMBFAST code in order to 
compute the CMB power spectrum under the presence of
scattering by atoms and ions. \\

\begin{flushleft}
\small {\it ANALYTICAL FORM OF $\delta C_l$'s:}
\end{flushleft}
Following eq.(\ref{eq:dT1})
for the {\it k}-mode of the photon temperature fluctuation,
\[
\Delta_T (k,\eta_0,\mu ) = \int_0^{\eta_0}d\eta \;e^{ 
ik\mu(\eta-\eta_0) } \ \  \]
\begin{equation}
\phantom{xxxxxxxxxxxxx}
\times \ 
\bigl[ \Upsilon (\eta ) \left( \Delta_{T0} 
 - i\mu v_b \right) +
\dot{\phi} + \psi  - ik\mu\psi \bigr], 
\label{eq:dT1b}
\end{equation}
we proceed now to characterize the change in the radiation power
spectrum. In
eq.(\ref{eq:dT1b}), the terms in the angle 
$\mu=\bf{\hat{k}}\cdot{\bf{\hat{n}}}$ can be eliminated after
integrating by
 parts and neglecting the contribution of boundary terms, 
(Seljak \& Zaldarriaga 1996). This gives:
\begin{equation}
\Delta_T = \int_0^{\eta_0}d\eta\; e^{ ik\mu(\eta -\eta_0)} S (k,\eta ),
\label{eq:dT2}
\end{equation}
where the source term $ S (k,\eta )$ is defined as
\begin{equation}
S(k,\eta ) = e^{-\tau}\left( \dot{\phi} + \dot{\psi}\right) +
        \Upsilon \left( \Delta_{T0} + \psi + \frac{\dot{v_b}}{k} \right)
 +      \dot{\Upsilon} \left( \frac{v_b}{k} \right).
\label{eq:s}
\end{equation}

From this source term, the angular power spectrum can be expressed as,
(e.g. Seljak \& Zaldarriaga, 1996):

\[
C_l  =  \left( 4\pi \right)^2
                \int dk \;k^2 P_{\psi} (k)\;
        \left| \int_0^{\eta_0} d\eta S(k,\eta ) 
                j_l \left[ k (\eta_0 - \eta) \right] \right|^2  \]
\begin{equation}
\phantom{xx}
        =  \left( 4\pi \right)^2
                \int dk \;k^2 P_{\psi} (k) \left| \Delta_{T,l}\right|^2,
\label{eq:cl}
\end{equation}
where $j_l$ is the spherical Bessel function of order $l$ and $P_{\psi}
(k)$ 
is the initial scalar perturbation power spectrum.
If at a given frequency the CMB interacts
through the  resonant transition $i$ of a species {\it X}, the measured
power 
spectrum will differ from the reference one by an amount:
\[
\delta C_l   \equiv   C_l^{X_i} - C_l \]
\[
\phantom{xx}
            =  ( 4\pi )^2
                \int dk\; k^2 P_{\psi} (k) \ 
         \bigl[  
        \left| \int_0^{\eta_0} d\eta \;S^{X_i}(k,\eta ) 
           j_l \left[ k \left(\eta_0 - \eta \right) \right] \right|^2 \]
\[
\phantom{xxxxxxxxxxxx}
                                 - \left| \int_0^{\eta_0} d\eta \;S(k,\eta ) 
                j_l \left[ k (\eta_0 - \eta) \right] \right|^2 \bigr] \;\]
\[
\phantom{xx}
                =   \left( 4\pi \right)^2
                \int dk\; k^2 P_{\psi} (k)  \bigl[ \; 
                        \left| \Delta_{T,l}^{X_i} \right|^2 - \left|
\Delta_{T,l} \right|^2
                        \bigr]  \; \]
\begin{equation}
        \phantom{xx}  =  \left( 4\pi \right)^2
                \int dk\; k^2 P_{\psi} (k) \bigl[ 2\;\Delta_{T,l} + 
              \delta \Delta_{T,l} \bigr] \delta \Delta_{T,l}
\label{eq:dif1}
\end{equation}

with $\delta \Delta_{T,l}\equiv \int_0^{\eta_0}d\eta\;\left(
S^{X_i}(k,\eta)-
S(k,\eta) \right) j_l\left[k(\eta_0-\eta )\right]$, and where the term
 $S^{X_i}(k,\eta )$ refers to the sources including the species $X$. 
 Note that the cross term $2\;\Delta_{T,l}\cdot \delta \Delta_{T,l}$ only
arises after computing the difference of the power spectra, i.e., it is not
present if one computes the power spectrum of the difference of
two maps obtained at different frequencies. This
cross term is precisely the responsible of having $\delta C_l$ linear in
$\tau_{x_i}$, and hence also linear in the abundance of the species $X_i$. 
This term also accounts for the correlation existing between the temperature
fluctuations generated during recombination and those generated during the
scattering with the species $X$. This non-zero correlation is 
essentially due to those $k$ modes corresponding to
lengths bigger than the distance separating the two events, i.e.,
recombination and scattering with $X$, (Hern\'andez-Monteagudo \& Sunyaev,
in preparation).
Let us
now model the differential opacity due to this transition as
$\dot{\tau_{X_i}} = \tau_{X_i} {\cal P}(\eta )$, where ${\cal P}(\eta )$
is a
 profile function of area unity, $\int_0^{\eta_0}d\eta '\; {\cal P}(\eta
')=1$.
We write the optical depth as $\tau_{tot}(\eta ) = \tau (\eta ) +
 \tau_{X_i}(\eta )$, where the last term refers to the optical depth due
to the
 transition $i$ of the {\it X} species. It can be 
expressed as  $\tau_{X_i}(\eta ) = \tau_{X_i} \; {\cal A} (\eta ) = 
\tau_{X_i} \int_{\eta}^{\eta_0} d\eta '\; {\cal P}(\eta ')$, with
 ${\cal A}(\eta )$ the
area function of the profile ${\cal P}(\eta )$. We will assume that the
profile peaks at $\eta=\eta_{X_i}$ and that $\eta_-$ and $\eta_+$ are
such 
that ${\cal P} (\eta ) \approx 0$ for $\eta < \eta_-$ and $\eta >
\eta_+$.
Hereafter, $\eta_{X_i}$ will be referred to as {\it transition epoch}
or {\it line epoch}.
 If we add this new term to the opacity, and assume that 
$\tau_{X_i} \ll 1$, then we can expand the term $\delta \Delta_{T,l}$ in
a
 power series of $\tau_{X_i}$. In this case, we obtain:

\[
\delta \Delta_{T,l}  \  = \  
         \tau_{X_i} \cdot       
        \biggl\{        -\int_0^{\eta_+} d\eta \;
        j_l\left[ k(\eta_0-\eta )\right] {\cal A} (\eta )  \]
\[ \phantom{xxxx}  \times \ \left[ 
        e^{-\tau}\left( \dot{ \phi} + \dot{ \psi} \right) +
\dot{ \tau}e^{-\tau}\left( \Delta_{T0} + \psi
+\frac{\dot{v_b}}{k}\right)
 + \; 
e^{-\tau }\left( \ddot{\tau} + \dot{\tau}^2 \right)
                                \frac{v_b}{k} \right]  \]
\[ \phantom{xxxxxxx}
         + \ \int_0^{\eta_+} d\eta\; j_l\left[ k(\eta_0-\eta )\right] 
        {\cal P}(\eta )  e^{-\tau}  \]
\[ \phantom{xxxxxxxxxxxxx}
        \times \ \left[
                \left(  \Delta_{T0} + \psi +\frac{\dot{v_b}}{k}\right) + 
                \left( \frac{1}{{\cal P}}\frac{d{\cal P}}{d\eta } +
2\dot{\tau}\right)
                        \frac{v_b}{k}
        \right]
                \biggr\}   \]
\[   \]
\[ \phantom{xxxx}
        + \  \ \tau_{X_i}^2  \cdot   \biggl\{ 
        \frac{1}{2} \int_0^{\eta_+} d\eta \;
        j_l\left[ k\left(\eta_0-\eta \right)\right] {\cal A}^2 (\eta )
        \]
\[
        \phantom{xxxxx} \times \ \biggl[ 
        e^{-\tau}\left( \dot{\phi} + \dot{\psi} \right) +
        \dot{\tau}e^{-\tau}\left( \Delta_{T0} + \psi
+\frac{\dot{v_b}}{k}\right) 
 + \ e^{-\tau}\left( \ddot{\tau} + \dot{\tau}^2 \right)
                                \frac{v_b}{k}\;\; \biggr] \]
\[ \phantom{xxxxxxxxxx}
    + \ \int_0^{\eta_+} d\eta\; j_l\left[ k\left( \eta_0-\eta
\right)\right] 
         e^{-\tau} \; \times \]
\[ \phantom{xxxxxxxxxxxxxxxxx} \biggl[ 
                -\left(  \Delta_{T0} + \psi + \frac{\dot{v_b}}{k} \right) {\cal
                A}(\eta ) {\cal P} (\eta ) \]
\[ \phantom{xxxxxxxxxxxxxxxxxxx} \times \ 
            \frac{v_b}{k} \left( {\cal P}^2(\eta ) +
                        {\cal A}(\eta ) \left( -\frac{d{\cal P}}{d\eta } -
2\dot{\tau}
                                     \right) \right) \biggr] \biggr\}  \]

\[ \phantom{xxxx} + \; 
\phantom{x}{\cal O} (\tau_{X_i}^3 )\;  \]
\begin{equation}
        \phantom{xxxx}  =  \tau_{X_i} \cdot  {\cal D}_1 +  
                \tau_{X_i}^2 \cdot  {\cal D}_2 +  {\cal O}( \tau_{X_i}^3 ).
\label{eq:dD}
\end{equation}

It is worth to note that ${\cal D}_1$ contains both the suppression of
intrinsic CMB anisotropies, and the generation of new fluctuations at
line epoch. This expansion allows us writing $\delta C_l$ 
as a power series of $\tau_{X_i}$ as well,
\[
\delta C_l  \; = 
\phantom{x}
 \tau_{X_i} \cdot (4\pi)^2 \int dk\; k^2 P_{\psi} (k) 
                \left[  {\cal D}_1\;2\Delta_{T,l} \right] \]
\[\phantom{xxxx}
+\;  \tau_{X_i}^2 \cdot (4\pi)^2 \int dk\; k^2 P_{\psi} (k) 
                \left[ {\cal D}_2\;2\Delta_{T,l} + {\cal D}_2^2
                                                                \right] \]
\[\phantom{xxxx}
+\;  {\cal O}( \tau_{X_i}^3 )\]
\begin{equation}
\phantom{xxxx}
         =  \; \tau_{X_i} \cdot {\cal C}_1 + \tau_{X_i}^2 \cdot {\cal C}_2 +
                        {\cal O}( \tau_{X_i}^3 )
\label{eq:dcl}
\end{equation}

\begin{figure*}
\begin{center}
\includegraphics[width=10cm,height=9.5cm]{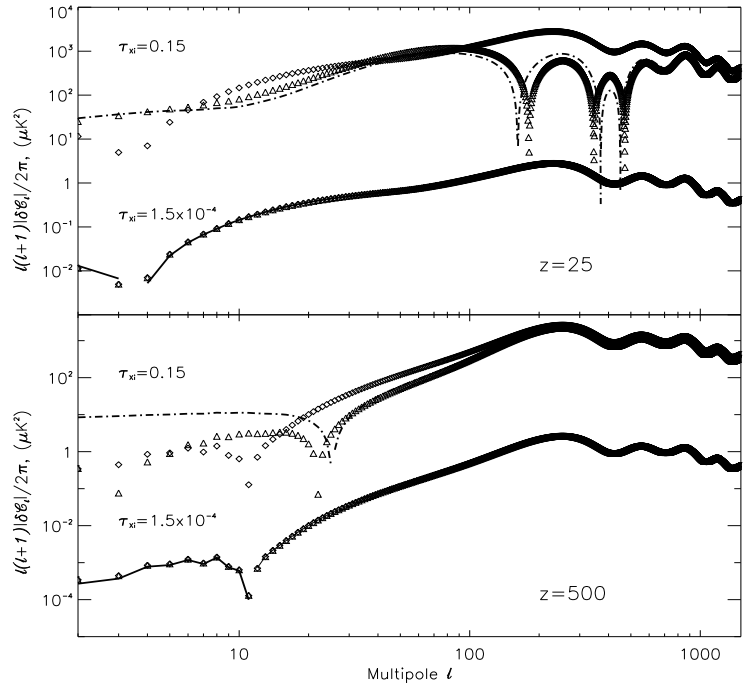}
\caption{ 
This figure shows the validity of the linear (diamonds) and quadratic
(triangles) approximations on $\tau$ when trying to describe the changes
induced in the power spectrum by a resonant transition ($\delta C_l$'s).
We see that both suffice to accurately match the theoretical  $\delta
C_l$'s
for redshifts 25 (top panel) and 500 (bottom panel) and for
$\tau=1.5\times
 10^{-4}$, (solid lines). However, when $\tau$ is closer to one
(dot-dashed
lines), the quadratic approximation performs remarkably better than the
linear one. 
}
\label{fig:l_cross}
\end{center}
\end{figure*}

\begin{figure*}
\begin{center}
\includegraphics[width=10cm,height=8.5cm]{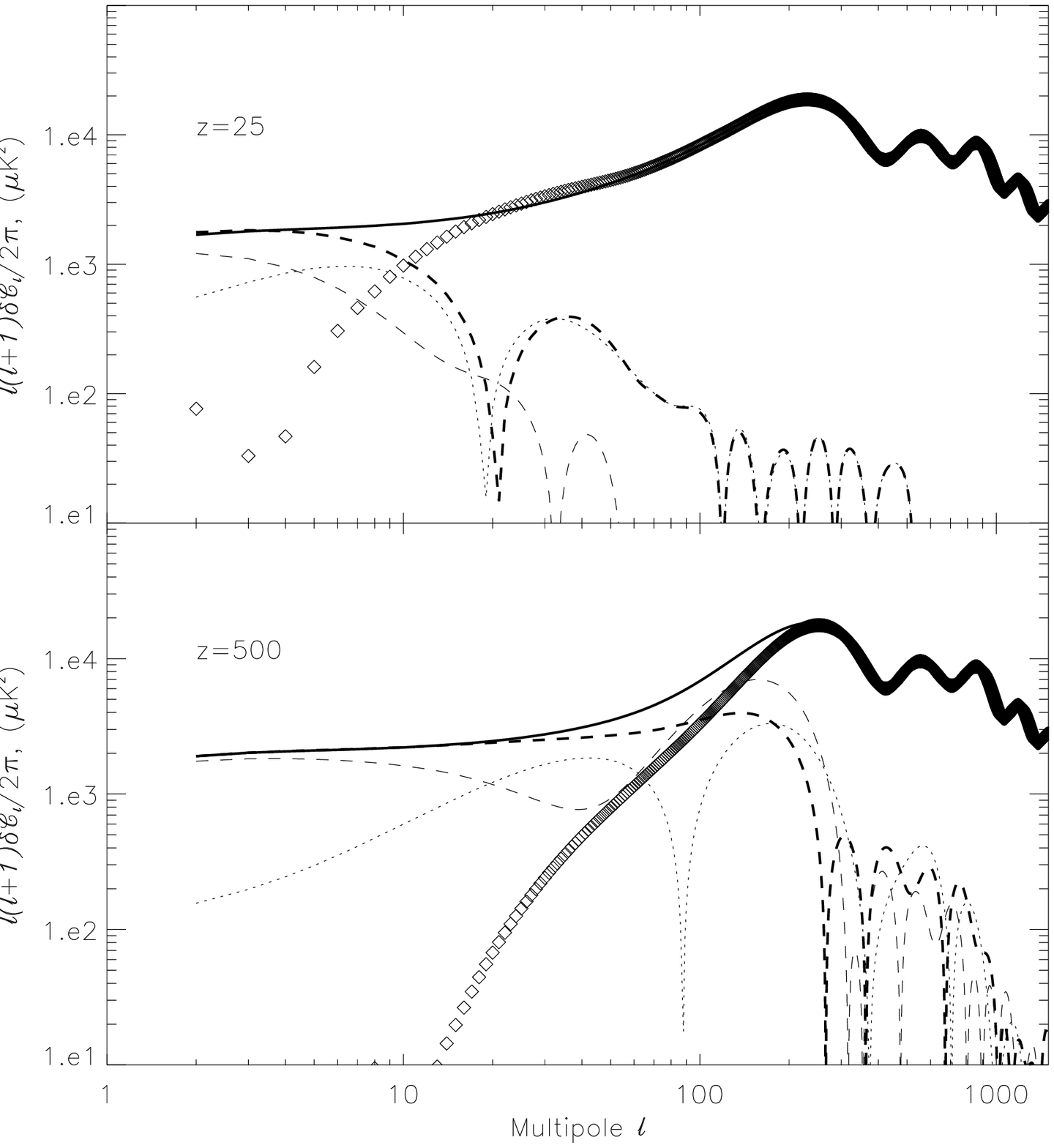}
\caption{ Study of the different contributions to ${\cal C}_1$ in 
eq.(\ref{eq:deltaCl}). Diamonds give ${\cal C}_1$ versus multipole.
We are plotting only absolute values.
 In units of $\tau_{X_i}$, we plot the damping term in solid line.
 The term responsible for the generation
of new anisotropies is plotted in thick dashed line. Its two components, 
monopole ($(\Delta_0 + \psi)$) and velocity, are also displayed in thin
dashed 
and thin dotted lines respectively. 
As conformal time goes by, the relative weight of the velocity
term with respect the monopole term increases, and the sum of both at
low
multipoles approaches the amplitude of the original CMB anisotropies,
(solid line). Both generation and damping tend to cancel each other at 
a multipole $l=l_{X_i}$, dependent on the epoch of interaction
$\eta_{X_i}$.}
\label{fig:terms}
\end{center}
\end{figure*}

The result of this expansion is displayed in Fig.\ref{fig:l_cross}: 
the change in the power
spectrum induced by resonant transitions is computed for different
redshifts, 25 (top panel) and 500 (bottom panel), and different
amplitudes of the optical depth (solid lines correspond to
$\tau=1.5\times 
10^{-4}$ and dot-dashed lines to $\tau=0.15$). Diamonds give the linear
approximation in $\tau$, whereas triangles show the quadratic one. Both
match the exact $\delta C_l$'s fairly well for the low $\tau$ cases.
Therefore, by means of eq.(\ref{sobolev}), we can establish 
a linear relation between $\delta C_l$ and the
abundance of the species.

 In Fig.\ref{fig:terms}, diamonds 
show the absolute value of ${\cal C}_1$
versus $l$. As shown above, this term is the sum of two integrals.
The first one is the suppression induced by the $e^{-\tau_{X_i}}$ term,
(thick
solid line in the figure), whereas
the second contain the newly generated anisotropies, (thick dashed
line).
The latter term has a monopole ($\Delta_0 + \psi$, thin dashed line) 
and a velocity (thin dotted line) contribution. 
We are plotting absolute values of all terms.
As pointed out by ZL02, the monopole term decreases
with cosmic time and hence the velocity term becomes the most important
source of generation of new anisotropies. However, this generation takes
place at the transition epoch, and hence is shifted towards the
low multipole range. It is easy to show that the maximum multipole
below which generation becomes significant is roughly given by
$l_{X_i} \sim 2\pi \; \left( \eta_0 - \eta_{X_i}\right)/\eta_{X_i}$. 
We show that the suppression term is dominant for
high multipoles, and only at very low multipoles suppression and
generation tend to cancel each other. 
Only if higher orders in the power expansion are
relevant, (i.e., if $\tau_{X_i} \sim 1$) the newly generated
anisotropies become important. 

This cancellation of suppression and generation terms 
at low multipoles can
be better understood when coming back onto eq.\ref{eq:dT1b}. If in
this equation we substitute $\tau$ by $\tau + \tau_{X_i}$, we find that
the change in the temperature modes due to $\tau_{X_i}$ can be written
as:

\[
\delta \Delta_T = \; \left( e^{-\tau_{X_i}} - 1 \right) \; 
        \int_0^{\eta_0} \; d\eta \; e^{ik\mu(\eta - \eta_0)}
        \dot{\tau} e^{-\tau} \left( \Delta_{T0} - i\mu v_b \right)
\]
\begin{equation}
\phantom{xxxxx}
          + \; e^{-\tau_{X_i}} \; 
        \int_0^{\eta_0} \; d\eta \; e^{ik\mu(\eta - \eta_0)}
        \dot{\tau}_{X_i} e^{-\tau} \left( \Delta_{T0} - i\mu v_b \right),
\label{eq:ddTap}
\end{equation}

where we have applied the definition of the visibility function,
$\Upsilon (\eta ) = \dot{\tau}(\eta) \; e^{-\tau(\eta)}$. However,
recalling that the visibility function gives the probability of a
photon being emitted at a given $\eta$, and taking $\dot{\tau}_{X_i} = 
\tau_{X_i}\delta_D (\eta - \eta_{X_i})$, the last equation merely
implies 
that
\begin{equation}
\delta \Delta_T \simeq \tau_{X_i}\left[
  \Delta_{T0}(\eta_{X_i}) - \Delta_{T0}(\eta_{rec})
                          - i\mu
                 \left( v_b(\eta_{X_i}) - v_b(\eta_{rec})
                        \right) \right],
\label{eq:ddTap2}
\end{equation}

where we have taken into account that $\tau_{X_i}$ is much smaller than
unity, and approximated the exponentials to unity, as we shall focus
on the very low $k$ range. 
That is, the change in the temperature mode $\Delta_T (k,\mu, \eta)$
reflects the difference of the monopole ($\Delta_{T0}$) and
velocity terms when evaluated at recombination and at the transition
epochs. We can estimate how this difference projects onto multipole
space if we restrict ourselves to the low multipole (large angle)
range.
For the scattering redshifts under
consideration ($z_{X_i} >$ a few), we can safely
neglect the term due to the Integrated Sachs-Wolfe effect.
Then from eq.(\ref{eq:dT1b}), one finds that the 
time dependence of the monopole term $\Delta_{T0}(\eta )$ can be
approximated as $\Delta_{T0}(\eta ) \simeq \Delta_{T0}(\eta_{rec} ) \;
j_0 \left( k \left[ \eta - \eta_{rec} \right] \right)$, where $j_0(x)$
is
the spherical Bessel function of order zero. From this dependence, it is
easy to see that at low multipoles, i.e., at small enough $k$'s,
$\Delta_{T0} (\eta_{rec} )$ and $\Delta_{T0} (\eta_{\eta_{X_i}} )$
will be roughly equal, and thus their difference in the equation above
will tend to vanish. This can also be seen in Fig.\ref{fig:terms},
where
the contribution of the monopole term to ${\cal C}_1$
(thin dashed line) has roughly
the same amplitude at redshifts 500 and 25 for $l < 10$. This behaviour
is the responsible for the cancellation of the $\delta C_l$'s at low
$l$'s.
In this situation, 
the difference of the velocity terms will be of particular relevance.
The 
evolution of velocities can be approximated, after integrating in $\mu$,
as $v_b (\eta) \simeq v_b (\eta_{rec}) \dot{D}(\eta ) /
\dot{D}(\eta_{rec} )j_1\left( k\left[ \eta - \eta_{rec}\right]\right)$, 
with $D(\eta)$ the
linear growth factor . The
growth of velocities will assure an increase of the Doppler-induced
anisotropy power. Finally, due to the fact that $j_l(x)$ is maximum at
$x\sim l$, then we have that the Doppler term will predominantly
contribute for 
\footnote{We are assuming
that $\eta_{X_i} >> \eta_{rec}$ throughout the paper. However, strictly
speaking, one should consider the difference of conformal times of
recombination and resonant scattering, i.e., 
$k \sim 2\pi/(\eta_{X_i}-\eta_{rec})$.} $k \sim 2\pi/\eta_{X_i}$, 
which corresponds to
multipoles 
$l_{X_i}\sim 2\pi \left(\eta_0 - \eta_{x_i}\right)/ \eta_{X_i}$.
Again, this can be checked by looking at the velocity
term (dotted line) in Fig.\ref{fig:terms}: the power is projected at
lower
 multipoles at later epochs, scales at which the amplitude grows 
with conformal time as $\dot{D}$.
\\

Therefore, the factors determining  the cross-over of the $\delta C_l$'s
from
(Doppler-induced) positive values at small $l$ 
to (absorption-induced) negative values at large $l$ are two:
{\it a)} the constant amplitude of the monopole at large scales, and
{\it b)}
the growth of peculiar velocities with cosmic time .
The angular scale at which such crossing takes place is roughly
determined by the time at scattering epoch,
$\eta_{X_i}$, $l_{X_i} \sim 2\pi \left(\eta_0 - \eta_{x_i}\right)/
\eta_{X_i}$. \\



\section[]{}

\begin{flushleft}
\small {\it LINE EMISSIVITY:}
\end{flushleft}
The volume emissivity for line emission due to collisional excitation
can be written as 

\begin{equation}
j_{\nu}^{line} = {\frac{h \nu}{4 \pi}} \ \langle n_i \ n_c \rangle 
\ \gamma_{lu} \ \varphi (\nu'),
\label{ap_eq_1}
\end{equation}

expressed in $erg \ cm^{-3} s^{-1} Hz^{-1} str^{-1}$ in the rest frame
of the emitter. Here $n_i$ is the number density of the atom or ion 
 under consideration, and $n_c$ is the number density of the 
colliding particle, and $\langle n_i\ n_c \rangle$ 
means averaging over the volume
specified by multipole $l$, corresponding to the angular size of
observation, and the thickness of the slice along the line-of-sight
defined by the frequency resolution of the experiment. 
$h \nu$ is the energy of the emitted photon,  
$\gamma_{lu}$ is the collision rate-coefficient
between between the two levels, and $\varphi (\nu')$ is the
line-profile function, which we can approximate by a $\delta$-function
because experimental bandwidth is much larger than velocity or thermal
broadening, 

\begin{equation}
\varphi (\nu') = \delta (\nu' - \nu) = {\frac{1+z}{\nu}} \ 
\delta (z' - z)\ , 
\label{ap_eq_2}
\end{equation}

where $z$ corresponds to the redshift under
study. Eqn.(\ref{ap_eq_1}) is correct for the low density gas 
($n \ll n_{crit}$) that we are concerned with in this paper. This
emissivity produces a line flux, written as following in an expanding
universe (neglecting absorption because $\tau_{\nu} \ll 1$)

\begin{equation}
F_{\nu}^{line} = \int {\frac{j_{\nu}}{(1+z)^3}} \ dl,
\label{ap_eq_3}
\end{equation}

where the integration is along the line-of-sight. 
This creates a distortion in the thermal spectrum of the background CMB
photons

\begin{equation}
{\frac{\Delta T}{T_{\gamma}}} = {\frac{e^x-1}{x e^x}} \
{\frac{F_{\nu}}{B_{\nu}}} \ , \ \ \ \ \ \ \ \ \mbox{where} \ \ 
x \equiv {\frac{h \nu}{k T_{\gamma}}}
\label{ap_eq_4}
\end{equation}

with $T_{\gamma}$ as the mean temperature of CMB photons, 
and $B_{\nu}$ being the Planck function. \\

\begin{flushleft}
\small {\it EFFECT OF OVER-DENSITY:}
\end{flushleft}
In this subchapter, we consider the impact of collisions in our study.
Collisions with electrons (or neutral atoms in the case of neutral gas)
cause emission from the same atomic 
and ionic fine-structure lines and hence change the observed $\delta
C_l$-s by producing an additional and independent signal 
on the top of the scattering signal. 

In addition, collisions are decreasing the population of the ground level
in the transitions under consideration, dimishing the optical depth
due to scattering in the line. In what follows,
we compute the deviation of the equilibrium
excitation temperature of a two-level system from the background
cosmic microwave radiation temperature as a function of the
over-density of the enriched region. 
we estimate the required over-density for causing a given
amount (30\% or 50\%) decrease in the optical depth. 

We consider a two-level system (like CII or NIII ion) for
simplicity, whose population ratio between the two levels is
determined by the {\it excitation temperature}, $T_{EX}$, defined as

\begin{equation}
(n_u / n_l) = (g_u / g_l) \exp \left[ 
-h \nu / kT_{EX} \right]
\label{ap_eq_7}
\end{equation}

where $n_u$ and $n_l$ are the relative fraction of atoms/ions in 
{\it upper} or {\it lower} levels, respectively, and $g_u$ and $g_l$
are the statistical weights of each level. The equilibrium population
ratio is obtained by solving  the statistical balance equation:

\begin{equation}
n_u \left( A_{ul} + B_{ul} J_{\nu} + n_e \gamma_{ul} \right) \ = \ 
n_l \left( \frac{g_u}{g_l} B_{ul} J_{\nu} + 
\frac{g_u}{g_l} n_e \gamma_{ul} e^{-h \nu / kT_K} \right)
\label{ap_eq_8}
\end{equation}

Here $A_{ul}$ and $B_{ul}$ are the Einstein coefficients for
spontaneous and stimulated emission, and $J_{\nu}$ 
is the background radiation field of CMB photons,
$J_{\nu} \equiv (2h \nu^3/c^2) [\exp (h \nu/kT_{CMB}) -
1]^{-1}$. $T_{CMB}$ is the radiation temperature at redshift $z$,
defined by 
$T_{CMB} \equiv T_0(1+z)$, with $T_0=2.726$ K. \ $n_e$ is the electron  
density inside the object at redshift $z$ , connected to its over-density
$\delta$ by the following: $n_e(z) \equiv \bar{n}_e(z) \ (1+\delta)$, where 
$\bar{n}_e(z) = 2.18 \times 10^{-7} (1+z)^3$ cm$^{-3}$. \ 
$\gamma_{ul}$ is the collisional de-excitation rate (in
cm$^3$ s$^{-1}$), and $T_K$ is the kinetic temperature of the
electrons, which we assume to be roughly constant at $10^4$ K. Then
from eqn.(\ref{ap_eq_7}) we can express the over-density as a function
of excitation temperature:

\begin{equation}
1+\delta = \frac {
A_{ul} \left[ (1+\Gamma) e^{-h \nu/kT_{EX}} - \Gamma \right] }
   {\bar{n}_e(z)\ \gamma_{ul} \left[ e^{-h \nu/kT_K} - 
e^{-h \nu/kT_{EX}} \right] }
\label{ap_eq_9}
\end{equation}

where $\Gamma \equiv [\exp (h \nu/kT_{CMB}(z)) - 1]^{-1}$, and we have
used $B_{ul} = (c^2/2h \nu^3)A_{ul}$. Eqn.(\ref{ap_eq_9}) is exact,
which can be further simplified in the limit of low redshifts, when
the line is in the Wien part of the CMB spectrum ($T_{CMB}(z)< T_{EX}
\ll h\nu$)  

\begin{equation}
1+\delta = \frac {A_{ul} \left[ e^{-h \nu/kT_{EX}} - e^{-h \nu/kT_{CMB}(z)}
\right] }{\bar{n}_e(z)\ \gamma_{ul} \left[ 1 - \frac{h
\nu}{kT_K} - e^{-h \nu/kT_{EX}} \right] }
\label{ap_eq_10}
\end{equation}

At $n_e < A_{ul}/\gamma_{ul}$ collisions do not influence the population 
of the ground level because radiative transitions are faster than
excitations due to collisions. 
If density is higher, population of the ground level  
decreases and optical depth due to resonance scattering
drops correspondingly, diminishing the effect we 
have discussed in this paper.

\begin{figure}
\includegraphics[width=8.3cm, height=5cm]{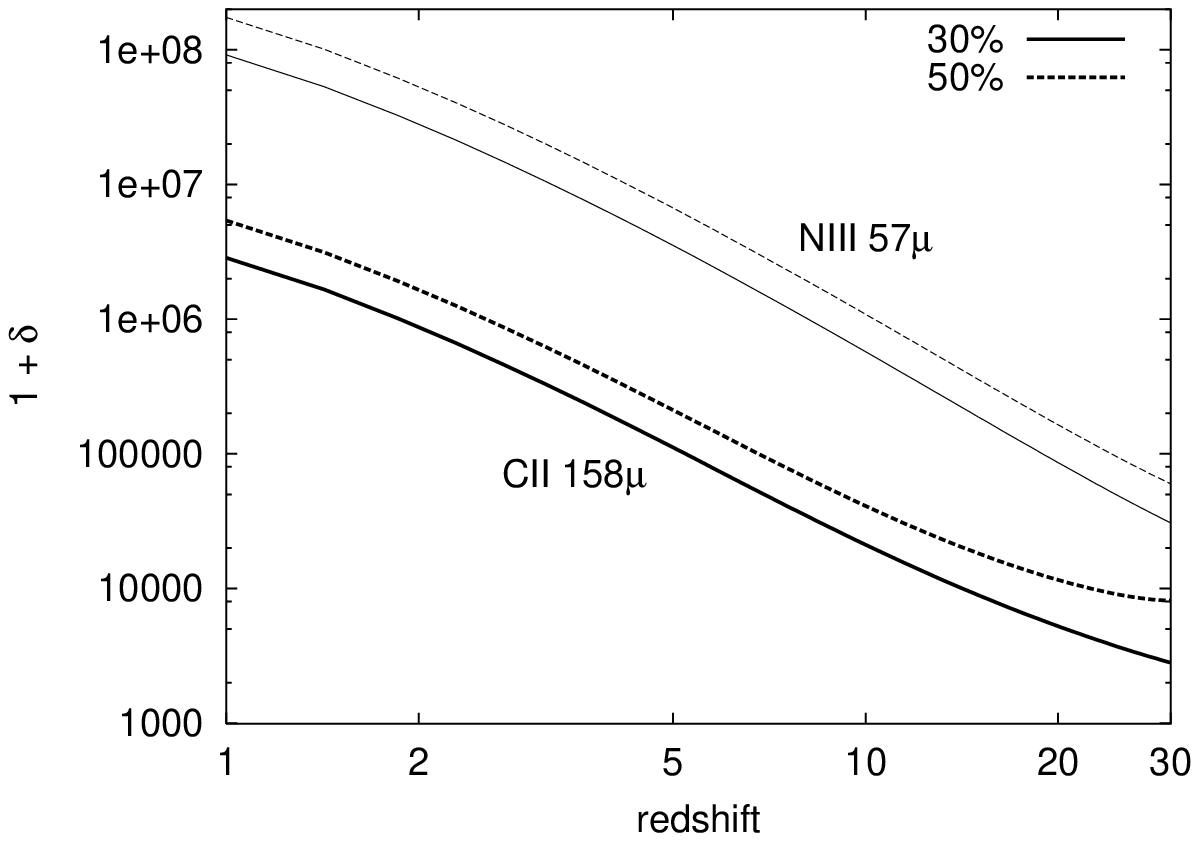}
\hspace*{.2cm}
\includegraphics[width=8.0cm, height=5cm]{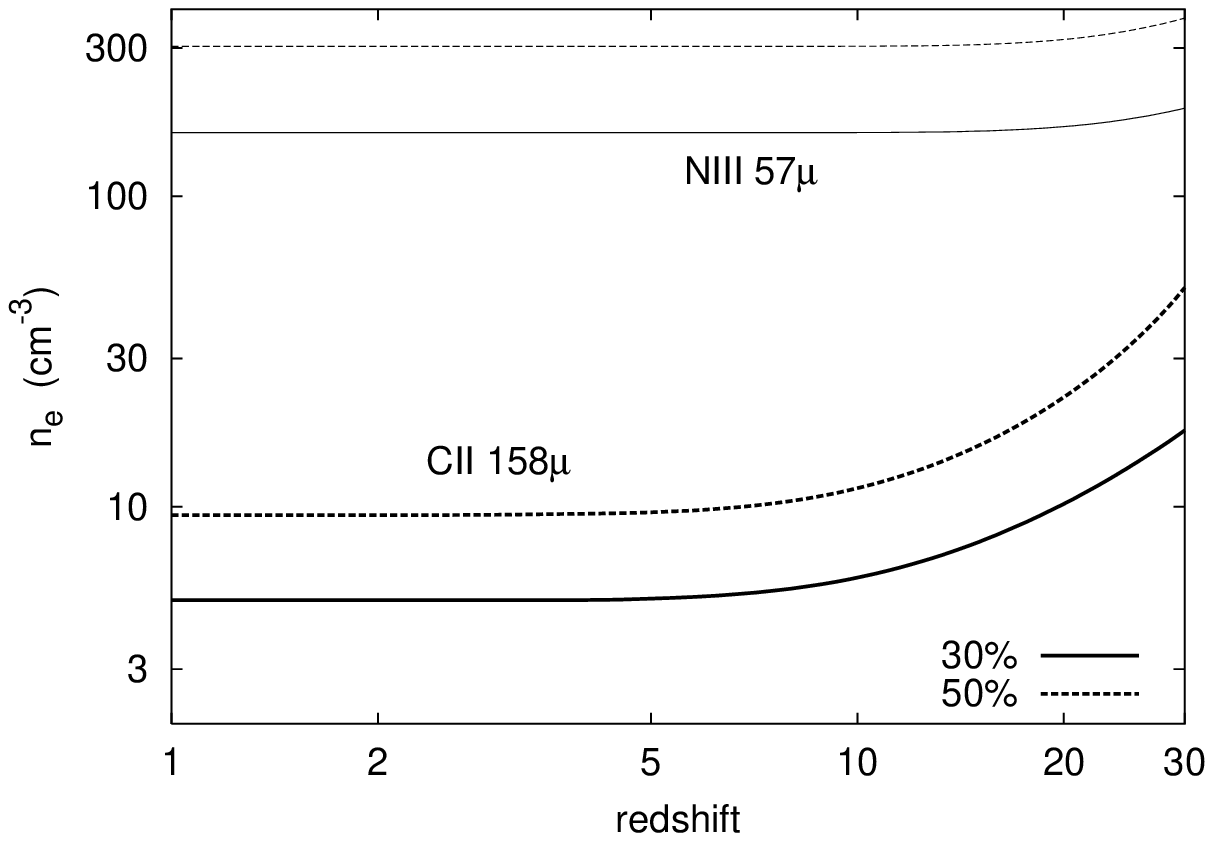}

\caption{Over-density and actual electron number density 
required to cause significant deviation of the
excitation temperature from the background cosmic microwave
temperature, as function of redshift. Shown here are the cases for CII
and NIII fine-structure singlets. The thick lines are for CII ion,and
the thin ones are for NIII ions. 
The pair of solid and dashed lines correspond to 30\%
and 50\% decrease in the optical depth for the same column density,
 respectively}
\label{ap_fig_1}
\end{figure}

Obviously, a very
large over-density is needed in order to produce any significant
amount of deviation from the CMB equilibrium, as can be seen from
Fig.\ref{ap_fig_1} for the case of CII 158$\mu$ and NIII 57$\mu$ 
transitions. These two transitions are picked as they both
arise from 
two-level fine-structure splitting, but results are similar for all
other relevant atoms and ions.  
In the bottom panel of Fig.\ref{ap_fig_1}, we see that the
electron density required to cause a given amount of depopulation of
the ground level is constant for low redshifts, and this density is
less than the critical density for that transition (e.g., with CII
158$\mu$ line, we have $\gamma_{ul}=4.6\times 10^{-8}$ cm$^3$ s$^{-1}$
for collision with electrons at $10^4$K,  
and the ratio $A_{ul}/\gamma_{ul}=50$ cm$^{-3}$. 
This simple estimate gives 5-10 times higher 
critical density than the more exact approach shown in
Fig.\ref{ap_fig_1}).  
Hence, we conclude that the effectiveness of the resonant scattering process
decreases in the most over-dense regions of the universe. 
Atoms and ions of heavy 
elements everywhere else including less over-dense halos and clouds
participate in the process and contribute into the
angular fluctuations of the CMB we discussed in this paper.

\end{document}